\documentclass[lettersize,journal]{IEEEtran}
\usepackage{cite}
\usepackage{caption}
\usepackage{subcaption}
\usepackage{float} 
\usepackage{graphicx}
\usepackage{amssymb}
\usepackage{amsthm}
\usepackage{pifont}
\usepackage{xcolor}
\usepackage[linesnumbered,ruled,vlined]{algorithm2e}
\usepackage{algorithmic}

\usepackage{amsmath,amsfonts}

\usepackage{array}
\usepackage{textcomp}
\usepackage{stfloats}
\usepackage{url}
\usepackage{verbatim}
\usepackage{graphicx}
\usepackage{cite}
\begin{document}
	
\newpage
\onecolumn
\centerline{\Large This work has been submitted to the IEEE for possible publication.}

\centerline{\Large Copyright may be transferred without notice, after which this version may no longer be accessible.}

\twocolumn
\title{Slice-Level Scheduling for High-Throughput and Load-Balanced LLM Serving}

\author{Ke Cheng, Wen Hu, Zhi Wang, Hongen Peng, \\
		Jianguo Li~\IEEEmembership{Senior~Member,IEEE,}
	    Sheng Zhang~\IEEEmembership{Senior~Member,IEEE}

\thanks{Ke Cheng and Sheng Zhang are with the State Key Laboratory for Novel Software Technology, Nanjing University, Nanjing 210023, China	
	(e-mail: ketonmi@outlook.com, sheng@nju.edu.cn).}% <-this % stops a space
\thanks{Wen Hu, Zhi Wang, Hongen Peng, and Jianguo Li are with Ant Group, Hangzhou 310063, China (e-mail: \{huwen.hu, wangchun.wz, hongen.phe, lijg.zero\}@antgroup.com).}
\thanks{This work was done during Ke Cheng's internship at Ant Group.}

\vspace{-0.5cm}
}



\maketitle

\begin{abstract}
Large language models (LLMs) iteratively generate text token by token, with memory usage increasing with the length of generated token sequences. Since the request generation length is generally unpredictable, it is difficult to estimate the time and memory required to process requests, thus posing a challenge for effective request scheduling. Conventional sequence-level scheduling (SLS) serves requests in a first-come first-served (FCFS) manner with static batching where requests with short generation lengths are delayed until those with long ones have finished generation. Besides, to avoid out-of-memory (OOM) errors, SLS batches requests using a small batch size, which limits throughput. Recently proposed iteration-level scheduling (ILS) improves this with continuous batching, timely completing requests and dynamically adding new ones, but often limits the number of parallel-processing requests to OOM errors, thus compromising throughput. Moreover, both SLS and ILS fail to effectively balance workload across multiple LLM instances.
To tackle these challenges, we propose slice-level scheduling (SCLS).
By splitting the predefined maximal generation length limit into slices and serving batches slice by slice, it provides a precise range of serving time and memory usage for batched requests, laying the foundation for effective scheduling. Experiments confirm that compared with SLS and ILS schedulers, SCLS can improve throughput by up to 315.8\% and greatly mitigate load imbalance with proposed batching and offloading algorithms. 
\end{abstract}


\section{Introduction}
\IEEEPARstart{L}{arge} language models (LLMs) are currently playing an important role in a wide range of applications, especially interactive web chatbots, such as CodeFuse \cite{codefuse} developed by Ant Group. In addition to the daily chat capability, CodeFuse is also widely acclaimed for its enhanced ability to generate, optimize, and interpret various programming languages, including SQL, Java, Python, and C++, boasting a peak concurrent usage with thousands of online users. Therefore, an efficient serving system with high throughput and scalability is essential to handle such a heavy workload.

To achieve high throughput for LLM serving, it is significant to fully utilize the parallel computing capability of GPUs with batch serving. Static batching and continuous batching are two batch-serving techniques for LLM inference. Static batching pads the raw input of batched requests to the same length before serving and iteratively generates tokens until all requests generate the end-of-sequence (EOS) token or the predefined iteration number limit is reached. Continuous batching leverages customized CUDA kernels to remove request padding and enable dynamic exits and joins of requests during batch serving. 

However, the time and GPU memory required for serving a request increases with the generation length (i.e., the number of generated tokens) because of the autoregressive generation pattern and key-value cache mechanism of LLM inference.
Since the request generation length is generally unpredictable, it is difficult to estimate the serving time and memory consumption of requests in advance. Therefore, it is challenging to schedule requests to take full advantage of these batch-serving techniques for high throughput.

\begin{figure*}[t]
	\setlength{\belowcaptionskip}{-0.05cm} 
	\centering
	\begin{subfigure}{0.35\linewidth}
		\centering
		\includegraphics[width=1\linewidth]{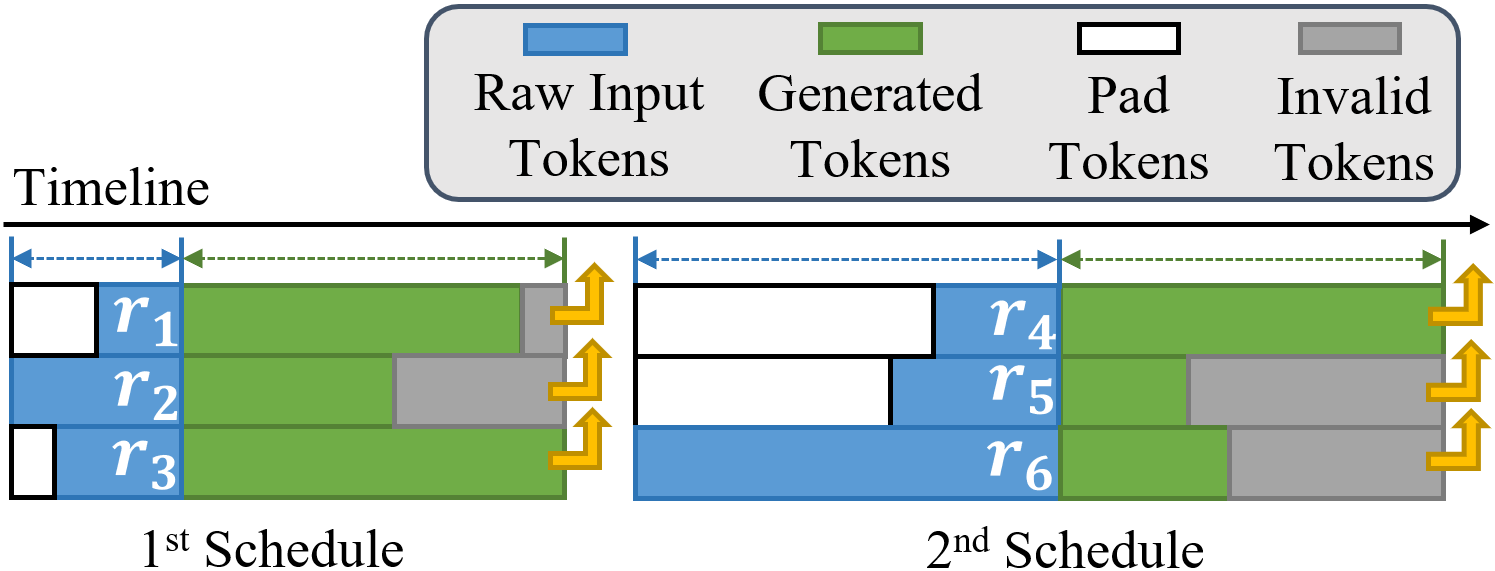}
		\caption{Sequence-level scheduling (SLS).}
		\label{fig:sls}
	\end{subfigure}
	\centering
	\begin{subfigure}{0.325\linewidth}
		\centering
		\includegraphics[width=1\linewidth]{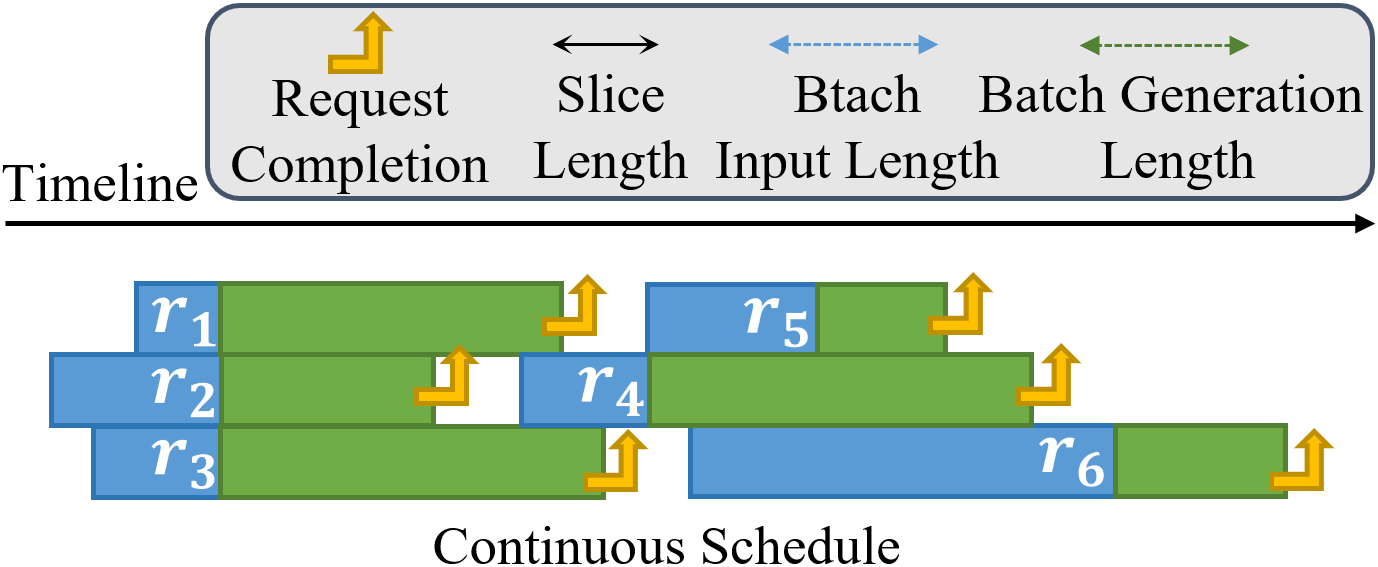}
		\caption{Iteration-level scheduling (ILS).}
		\label{fig:ils}
	\end{subfigure}
	\centering
	\begin{subfigure}{0.3\linewidth}
		\centering
		\includegraphics[width=1\linewidth]{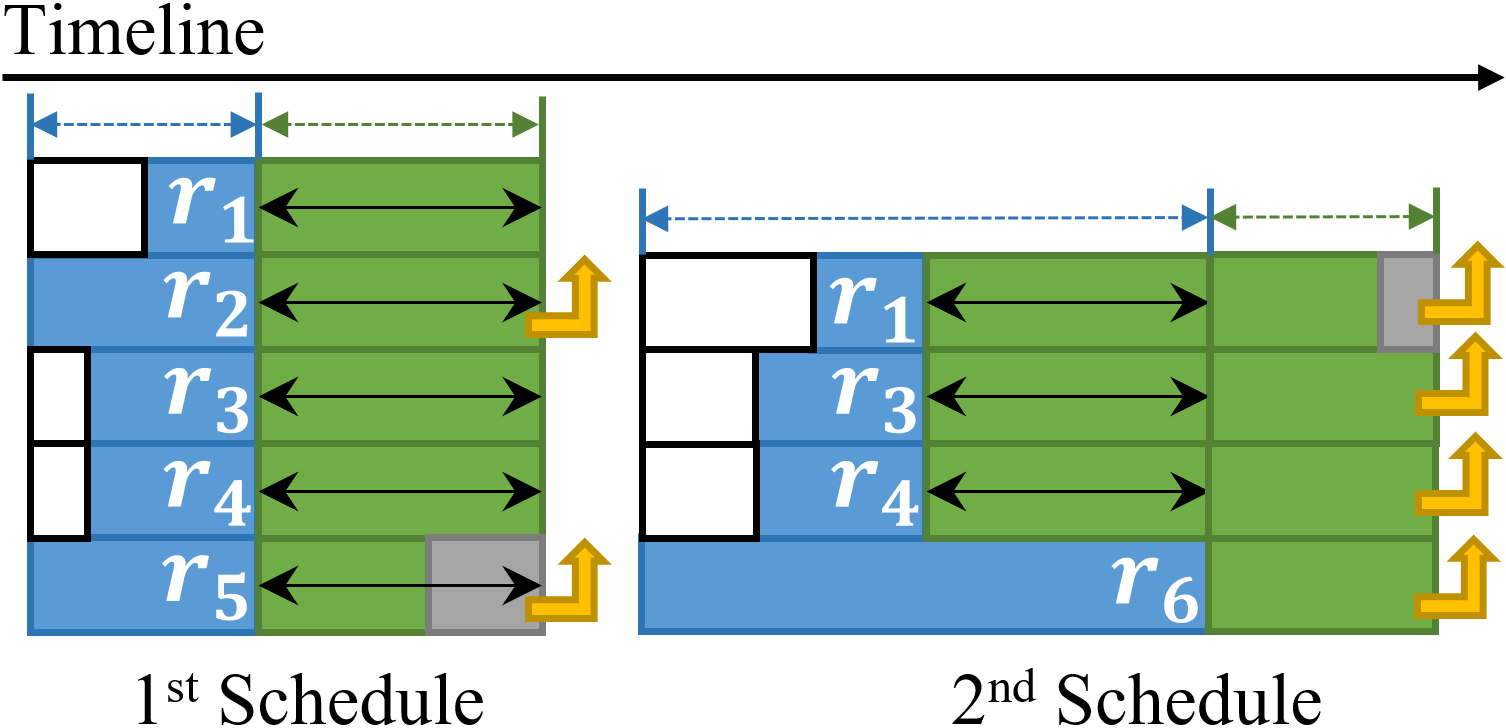}
		\caption{Slice-level scheduling (SCLS).}
		\label{fig:scls}
	\end{subfigure}
	\caption{Illustration of various scheduling techniques, where $r_i$ represents the $i$th arrived request.}
	\vspace{-0.5cm}
	\label{fig:various-scheduling-methods}
\end{figure*}

Existing deep learning serving systems \cite{triton, tensorflowserving} adopt the sequence-level scheduling (SLS) to batch requests in a first-come first-served (FCFS) manner with a fixed batch size and serve them with static batching as shown in Fig. \ref{fig:sls}, which leads to severe computational inefficiencies. Firstly, for static batching, in a batch, requests with short inputs are padded, and requests with short outputs have to wait for those with long outputs to finish before they are returned together. During the waiting, completed requests still participate in the computation and generate invalid tokens. Secondly, without knowing request generation lengths, the SLS scheduler directly sets the iteration number limit for static batching to the predefined maximal generation length and defaults to a small batch size to prevent out-of-memory (OOM) errors, thus failing to fully exploit the GPU's parallel computing capability and compromising throughput.

To eliminate the request padding and waiting caused by SLS and improve computational efficiency, iteration-level scheduling (ILS) is introduced in recent studies \cite{yu2022orca, kwon2023efficient, deepspeed-fastgen}. Current ILS techniques leverage continuous batching to timely return completed requests and add newly arrived requests for processing in an FCFS manner at each iteration, as shown in Fig. \ref{fig:ils}. 
However, without knowing request generation lengths, many ILS techniques \cite{yu2022orca, deepspeed-fastgen} adopt conservative memory management strategies that limit the number of parallel-processing requests to avoid OOM errors, thus under-utilizing the GPU and hurting throughput.

Moreover, existing SLS and ILS schedulers \cite{triton, tensorflowserving, deepspeed-fastgen} leverage the round-robin policy to offload requests to multiple LLM instances, which causes unbalanced distribution of computational overhead and memory consumption across LLM instances due to the variance of request generation lengths.

In this paper, we propose slice-level scheduling (SCLS) to address the aforementioned challenges from a pure scheduling perspective. The core idea is to split the predefined maximal generation length limit into fixed-length slices and serve batches slice by slice. Completed requests are returned promptly, while incomplete requests are rescheduled with newly arrived requests until finished. Generation slicing can enhance both static and continuous batching. We focus on strengthening static batching in this paper. In Section \ref{sec:discussion}, we discuss promising technique solutions for applying the idea of generation slicing to continuous batching, which we leave as our future work.

Compared with the maximal generation length limit, the small slice length results in a narrow range of generation length at each schedule, thus providing a precise range of memory consumption and serving time for batches. With the precise range of memory consumption, requests can be adaptively batched together with as large a batch size as possible to improve the throughput without exceeding the GPU memory as shown in Fig. \ref{fig:scls}. With the precise range of serving time, batches can be scheduled to LLM instances using the max-min policy \cite{radunovic2007unified} to achieve load-balancing. All in all, generation slicing lays the foundation for effective scheduling. Besides, since requests with long outputs are rare in real-world applications such as ChatGPT \cite{chatgpt} and CodeFuse, most requests can be returned in a few slices without waiting for long.

We implement SCLS as the gateway to route requests and conduct extensive experiments with LLM instances deployed in the production environment of Ant Group using CodeFuse request traces. Experimental results show that SCLS can achieve a high request throughput and a more balanced workload across LLM instances. Our main contribution is listed as follows.

\begin{itemize}
	\item We confirm the inefficiency and load imbalance of SLS and  ILS schedulers through experiments. Besides, we find that requests with long outputs are rare by analyzing the ShareGPT dataset and request traces collected from CodeFuse's logs. These motivate the birth of SCLS.
	\item We propose SCLS. By limiting the number of iterations to a small slice length in each schedule, it controls the serving time and memory usage of batches to a precise range, laying the foundation for effective scheduling.
	\item We design a dynamic programming-based batching algorithm and a max-min-based offloading algorithm for SCLS. They divide requests into batches to minimize the batch serving time and balance the computational workload across multiple LLM instances, respectively.
	\item We conduct extensive experiments to confirm the superiority of SCLS. Equipped with various inference engines, SCLS can improve request throughput by up to 315.8\%, reduce response time by up to 91.1\%, and greatly mitigate load imbalance compared with SLS and ILS schedulers. 
\end{itemize}

\section{Preliminaries}
LLMs generate text by predicting the next token for an input token sequence and adding the predicted token to the input iteratively until the EOS token is generated. Such a generation pattern is called \textbf{auto-regressive generation}. To prevent the generation process from being too long, a maximal generation length limit is always set. When this limit is reached, even if the EOS token has not been generated, the generated results are returned to users. The \textbf{request input length} is the number of tokens in the raw request text, while the \textbf{request generation length} is the number of generated tokens. 

In static batching, requests are padded to the same length and processed together. Therefore, the \textbf{batch input length} is the length of the longest request in the batch. The \textbf{batch generation length} is the total number of iterations for the batch, depending on the longest request generation length and the iteration number limit. 

To accelerate LLM inference, the key and value tensors produced by transformer blocks of LLMs are cached for reuse. In the first iteration of the auto-regressive generation, the token sequence of raw request text is fed to the LLM, and the key and value tensors of these input tokens are computed and cached. In subsequent iterations, only the latest generated token is fed to the LLM for computation and cache of key and value tensors. Thus, the generation process can be split into two phases. The first iteration is the computationally heavy \textbf{prefill phase}. Subsequent iterations are the computationally light \textbf{decoding phase}.

\begin{figure*}[t]
	\setlength{\belowcaptionskip}{-0.05cm} 
	\centering
	\begin{subfigure}{0.19\linewidth}
		\centering
		\includegraphics[width=1\linewidth]{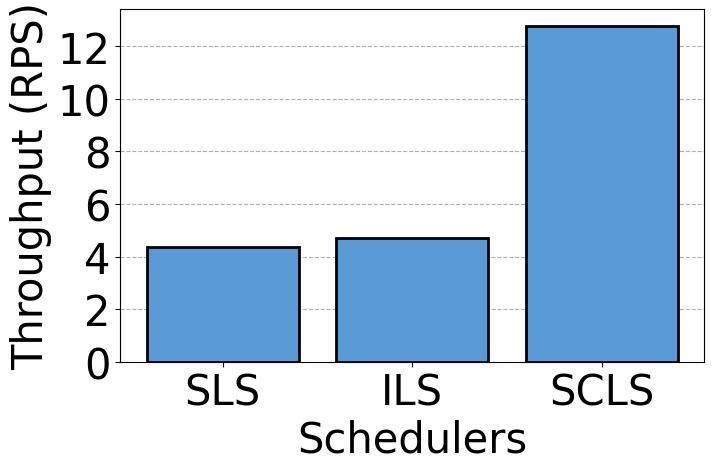}
		\caption{Throughput.}
		\label{fig:motivation_throughput}
	\end{subfigure}
	\centering
	\begin{subfigure}{0.19\linewidth}
		\centering
		\includegraphics[width=1\linewidth]{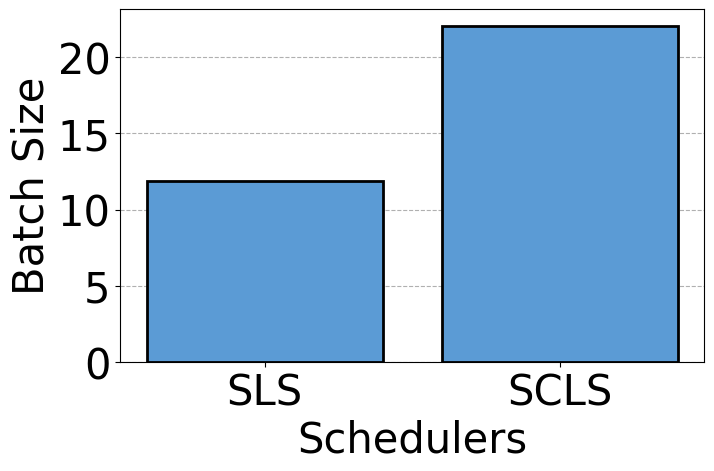}
		\caption{Batch size.}
		\label{fig:moivation_batch_size}
	\end{subfigure}
	\centering
	\begin{subfigure}{0.1955\linewidth}
		\centering
		\includegraphics[width=1\linewidth]{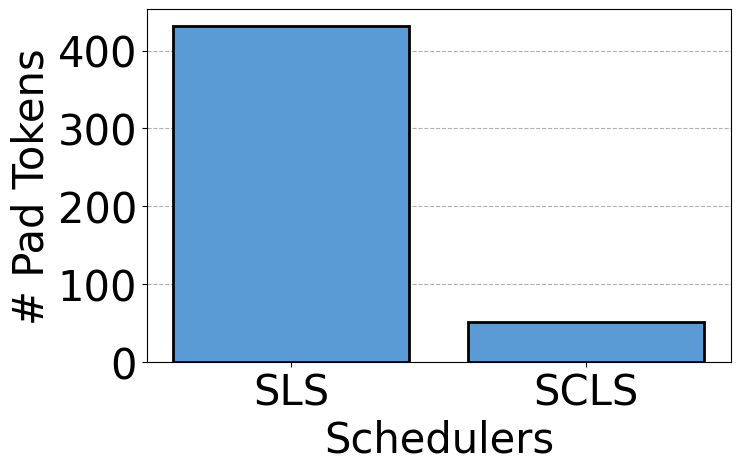}
		\caption{\#Pad token.}
		\label{fig:motivation_pad_tokens}
	\end{subfigure}
	\centering
	\begin{subfigure}{0.196\linewidth}
		\centering
		\includegraphics[width=1\linewidth]{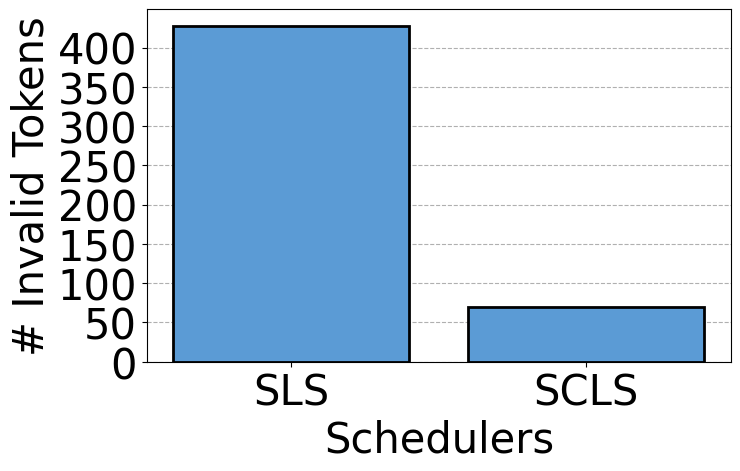}
		\caption{\#Invalid token.}
		\label{fig:motivation_invalid_tokens}
	\end{subfigure}
	\centering
	\begin{subfigure}{0.19\linewidth}
		\centering
		\includegraphics[width=1\linewidth]{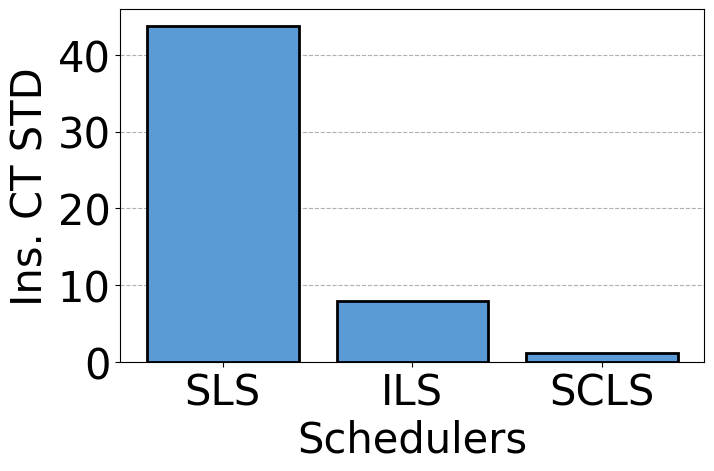}
		\caption{Load imbalance.}
		\label{fig:motivation_load_imbalance}
	\end{subfigure}
	\caption{Inefficiency and load imbalance of SLS and ILS schedulers. Sub-figures (a) and (e) show that SCLS can significantly outperform SLS and ILS schedulers in terms of throughput and load balancing. Sub-figures (b)\textasciitilde(d) present that SCLS's superior performance comes from the increase of batch size and decrease of the number of pad tokens and invalid tokens.}
	\vspace{-0.25cm}
	\label{fig:motivation}
\end{figure*}

\section{Motivation}

\subsection{Inefficient Serving of SLS and ILS}
\label{sec:inefficient-batch-serving-for-sls}
We conduct experiments to confirm the inefficient serving of SLS and ILS schedulers. We implement the SLS and SCLS schedulers using deepspeed-inference \cite{deepspeed-inference} as the inference engine and leverage deepspeed-fastgen \cite{deepspeed-fastgen} as the ILS scheduler. In the experiment, 8 LLaMA2-13B \cite{touvron2023llama2} instances are deployed on 8 NVIDIA A100 80GB GPUs, and requests with various input and generation lengths arrive at a rate of 20 per second. Other settings are the same as in Section \ref{sec:experiment}. The experimental results are shown in Fig. \ref{fig:motivation}.

From Fig. \ref{fig:motivation_throughput}, we can see that the throughput of SLS is much smaller than that of SCLS. This is because SLS serves requests with a fixed small batch size. Moreover, SLS simply batches requests in an FCFS manner, which leads to a lot of pad tokens and generated invalid tokens caused by the variance of the batched requests' input lengths and generation lengths. 
Besides, although continuous batching can eliminate pad tokens and invalid tokens in the serving process, the throughput of ILS is also much lower than that of SCLS, because the ILS scheduler adopts a conservative memory management mechanism that limits the number of parallel-processing requests.

\begin{figure}[t]
	\setlength{\belowcaptionskip}{-0.05cm}
	\centering
	\begin{subfigure}{0.4925\linewidth}
		\centering
		\includegraphics[width=1\linewidth]{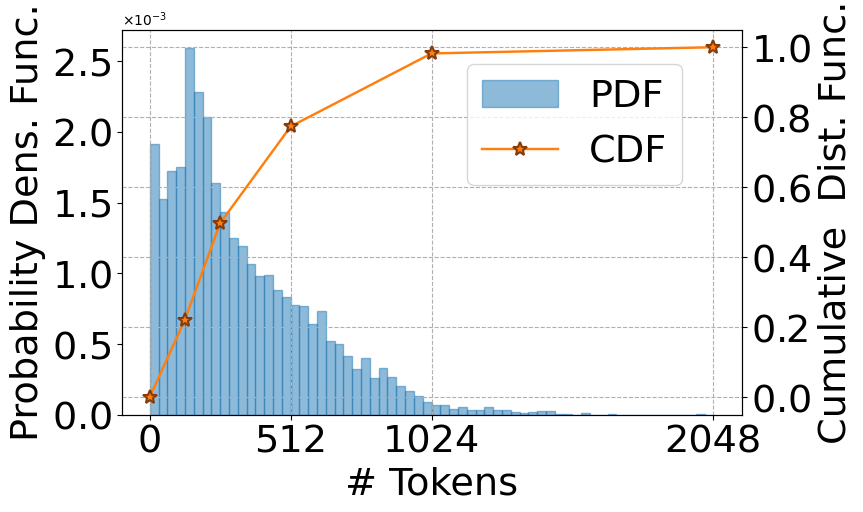}
		\caption{CodeFuse.}
		\label{fig:codefuse}
	\end{subfigure}
	\centering
	\begin{subfigure}{0.4925\linewidth}
		\centering
		\includegraphics[width=1\linewidth]{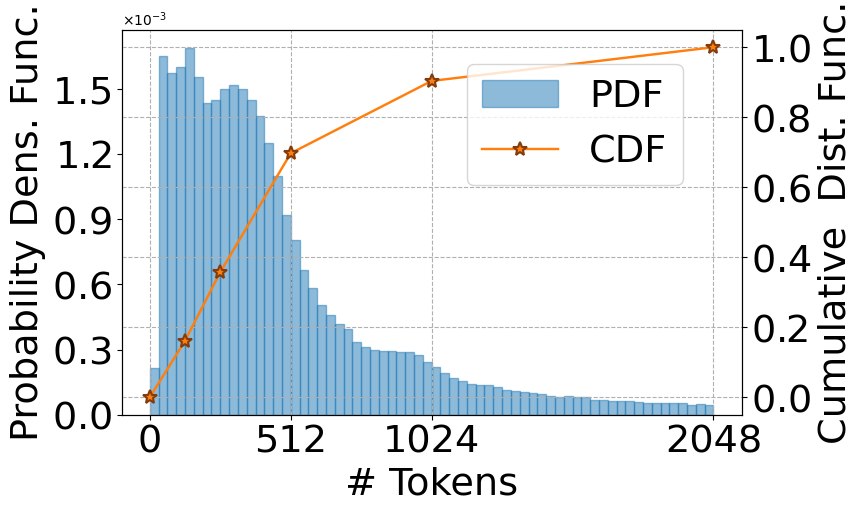}
		\caption{ShareGPT.}
		\label{fig:sharegpt}
	\end{subfigure}
	\caption{
		Probability density function (PDF) and cumulative distribution function (CDF) of the request generation length in actual scenarios.
	}
	\vspace{-0.25cm}
	\label{fig:output-length-distribution}
\end{figure}

\subsection{Unbalanced Workload of SLS and ILS}
We record the time when each LLM instance completes serving at the end of the experiment. Fig. \ref{fig:motivation_load_imbalance} presents the standard deviation (STD) of the completion time (CT) of LLM instances. Since both SLS and ILS schedulers use the round-robin policy to offload requests, requests with long and short generation lengths may be assigned to different instances, resulting in unbalanced memory consumption and computational overhead across instances. For the SLS scheduler, instances that have more requests with long generation lengths take more time to serve requests. For the ILS scheduler, instances that have more requests with long generation lengths consume more memory and cannot promptly add new requests for processing due to insufficient memory, causing a long request queuing time. Such a load imbalance accumulates to result in a large variance in the CT of LLM instances. Hence, the CT STD for LLM instances under SLS and ILS schedulers is much larger than that of SCLS.

\begin{figure}[t]
	\setlength{\belowcaptionskip}{-0.05cm} 
	\centering
	\includegraphics[width=1\linewidth]{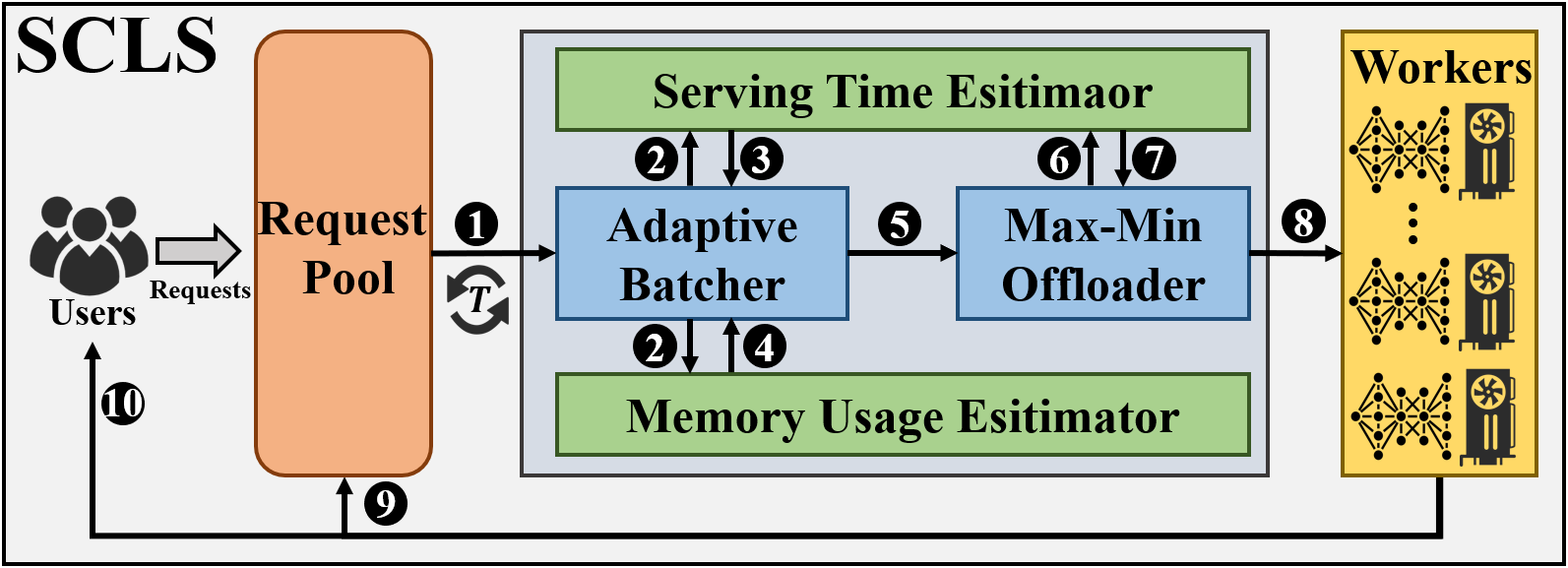}
	\caption{SCLS system overview. The arrows represent data transferred between modules, where \ding{182} is requests, \ding{183} is temporary batches, \ding{184} and \ding{188} are estimated serving time, \ding{185} is estimated memory consumption, \ding{186}, \ding{187}, and \ding{189} are batches, \ding{190} is uncompleted requests, and \ding{191} is completed requests.}
	\vspace{-0.25cm}
	\label{fig:system_overview}
\end{figure}

\subsection{Real-World Generation Length Distribution}
To understand the generation length distribution in real-world LLM-based web applications, we analyze 400,000 pieces of data from ShareGPT\footnote{ShareGPT \cite{sharegpt} is a publicly available website that shares conversations from users chatting with ChatGPT \cite{chatgpt}.} and one-month request traces collected from CodeFuse's logs. We depict the distribution of request input and generation lengths in Fig. \ref{fig:output-length-distribution}, where we can find that the vast majority of requests have a small generation length of less than 512 in both CodeFuse and ShareGPT data. Hence, if we split the maximal generation length limit (e.g., 2048) into slices with a small slice length (e.g., 128) and serve batched requests slice by slice, most requests can be returned in time within a small number of slices. In addition, since the iteration number limit is set to the small slice length, we can get a precise range of memory consumption and batch together as many requests as possible while ensuring that OOM errors will not occur, thus increasing throughput.

Moreover, for static batching, we can also get a precise range of batch serving time. Hence, load balancing can be achieved by offloading batches with the max-min policy where the batch with the longest serving time is offloaded to the least loaded instance. 

Considering the above, we propose SCLS, which splits the predefined maximal generation length limit into fixed-length slices, limits the number of iterations in each batch serving to the slice length, and serves requests slice by slice. Although requests with long generation lengths suffer from the extra overhead of recomputing the prefill phase at every reschedule, the overhead is acceptable because these long-generated requests are rare as shown in Fig. \ref{fig:output-length-distribution}.

\section{Solution Description}
\label{sec:solution}

\subsection{System Overview}
\label{sec:solution-system-overview}
SCLS has four core components, a serving time estimator, a memory usage estimator, an adaptive batcher, and a max-min offloader. The overview of SCLS is presented in Fig. \ref{fig:system_overview}. 

SCLS periodically fetches all requests from the request pool with a time interval $T$ and hands them over to the adaptive batcher. 
The adaptive batcher leverages a dynamic programming-based algorithm to group the requests into batches with the goal of minimizing the total serving time. In each step of the batching algorithm, the adaptive batcher leverages the estimated serving time of temporarily grouped batches to update the state transfer function and utilizes estimated memory consumption to guarantee that the serving process will not cause OOM errors.
After the batching procedure finishes, batches will be offloaded to workers by the max-min offloader. After batch serving is finished, completed requests are returned to the user, and uncompleted requests are sent to the request pool to be scheduled with newly arrived requests.

A worker in Fig. \ref{fig:system_overview} is an LLM instance running on a specific inference engine. The worker's receiving thread receives the batch from the offloader and saves it in a local queue, while its processing thread continuously fetches a batch from the queue and serves it. In this paper, we implement the HF-worker and DS-worker to load and run LLM instances using huggingface transformers (HF) \cite{huggingface-transformers} and deepspeed-inference (DS) \cite{deepspeed-inference} as inference engines, respectively.

\begin{figure}[t]
	\setlength{\belowcaptionskip}{-0.05cm}
	\centering
	\begin{subfigure}{0.4975\linewidth}
		\centering
		\includegraphics[width=1\linewidth]{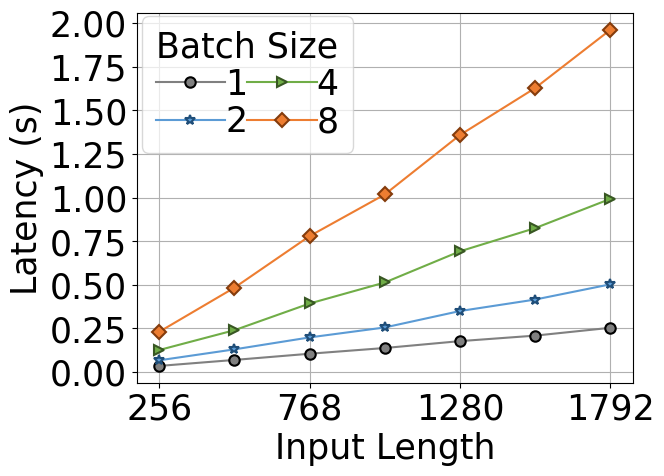}
		\caption{$T_{prefill}(N, L_i)$ under different input length.}
		\label{fig:prefill-input-length}
	\end{subfigure}
	\centering
	\begin{subfigure}{0.485\linewidth}
		\centering
		\includegraphics[width=1\linewidth]{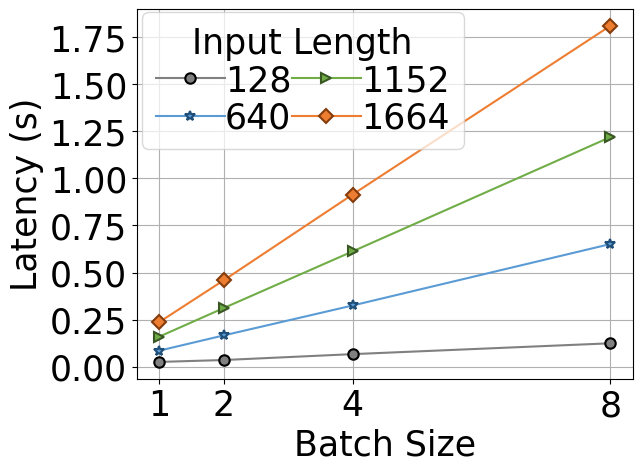}
		\caption{$T_{prefill}(N, L_i)$ under different batch sizes.}
		\label{fig:prefill-batch-size}
	\end{subfigure}
	\caption{Prefill latency under deepspeed-inference (DS).}
	\vspace{-0.15cm}
	\label{fig:prefill-latency}
\end{figure}

\begin{figure}[t]
	\setlength{\belowcaptionskip}{-0.05cm}
	\centering
	\begin{subfigure}{0.4975\linewidth}
		\centering
		\includegraphics[width=1\linewidth]{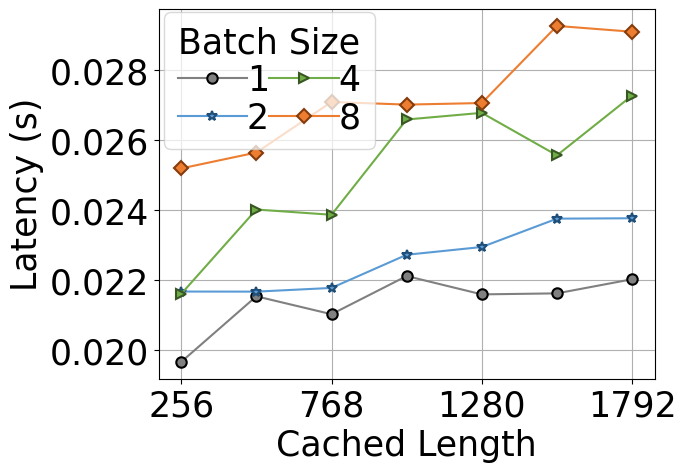}
		\caption{$\tau_{decode}(l, N)$ under different context lengths.}
		\label{fig:decode-input-length}
	\end{subfigure}
	\centering
	\begin{subfigure}{0.485\linewidth}
		\centering
		\includegraphics[width=1\linewidth]{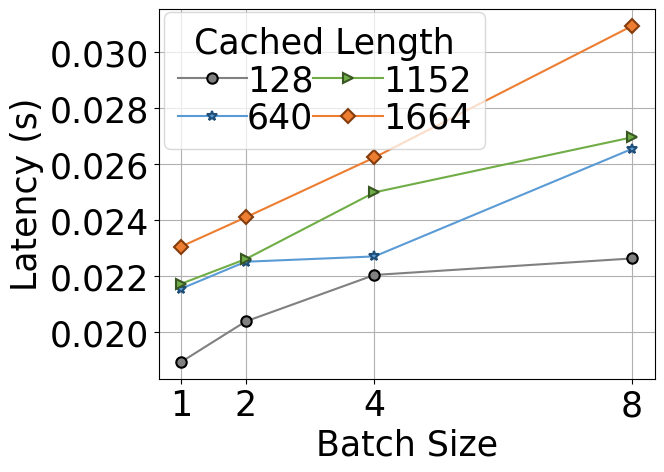}
		\caption{$\tau_{decode}(l, N)$ under different batch sizes.}
		\label{fig:decode-batch-size}
	\end{subfigure}
	\caption{Per-iteration decoding latency under DS.}
	\vspace{-0.25cm}
	\label{fig:decode-latency}
\end{figure}

\subsection{Efficient Serving Time Estimation}
\label{sec:solution-time-estimation}
Inference engines, such as HF and DS, serve requests with static batching, where the batch serving time is determined by the batch size, batch input length, and batch generation length.

Given a batch $\mathcal{B}$ whose batch size, batch input length, and batch generation length are separately represented as $N$, $L_i$, and $L_o$. The batch serving time is calculated by
\begin{equation}
	\label{eq:batch-serving-time}
	T_{serve}(N, L_i, L_o)= T_{prefill}(N, L_i) + T_{decode}(N, L_i, L_o),
\end{equation}
where $T_{prefill}(N, L_i)$, and $T_{decode}(N, L_i, L_o)$ are the serving latency in the prefill and decoding phase, respectively. Since the prefill phase computes key and value tensors for all the input tokens, the latency is determined by the batch input length $L_i$ and the batch size $N$. 

In the decoding phase, the serving procedure runs iteratively. Therefore, $T_{decode}(N, L_i, L_o)$ can be calculated by summing up the per-iteration decoding latency, which is expressed by 
\begin{equation}
	\label{eq:decoding-time}
	T_{decode}(N, L_i, L_o)=\sum_{l=L_i+1}^{Li+L_o} \tau_{decode}(l, N),
\end{equation}
where $\tau_{decode}(l, N)$ denotes the per-iteration latency. In each decoding iteration, for each request, only one token is fed to the LLM and extracts features from all the previously cached tokens to make a prediction, and hence $\tau_{decode}(l, N)$ is determined by the context length $l$ and the batch size $N$.

We profile the prefill latency $T_{prefill}(N, L_i)$ of an LLaMA2-13B instance using DS as the inference engine under various batch sizes and batch input lengths. To clearly show the individual effect of batch size $N$ and batch input length $L_i$ on $T_{prefill}(N, L_i)$, we plot two sub-figures with these two factors as the horizontal axis in Fig. \ref{fig:prefill-latency}. Similarly, we also profile the per-iteration latency $\tau_{decode}(l, N)$ and show the effect of context length $l$ and batch size $N$ in Fig. \ref{fig:decode-latency}.

\begin{figure}[t]
	\setlength{\belowcaptionskip}{-0.05cm}
	\centering
	\begin{subfigure}{0.5\linewidth}
		\centering
		\includegraphics[width=1\linewidth]{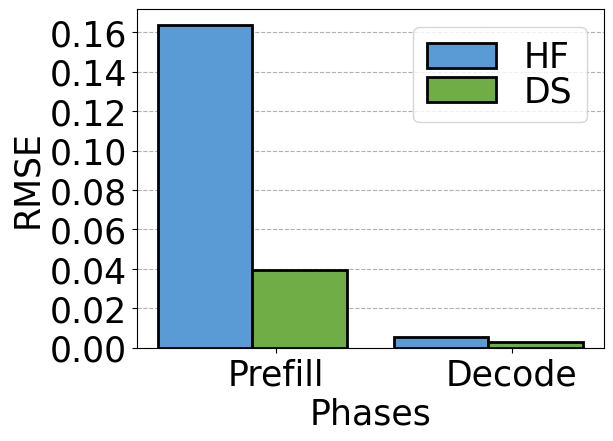}
		\caption{Error of 1 iteration.}
		\label{fig:error-iteration}
	\end{subfigure}
	\centering
	\begin{subfigure}{0.48\linewidth}
		\centering
		\includegraphics[width=1\linewidth]{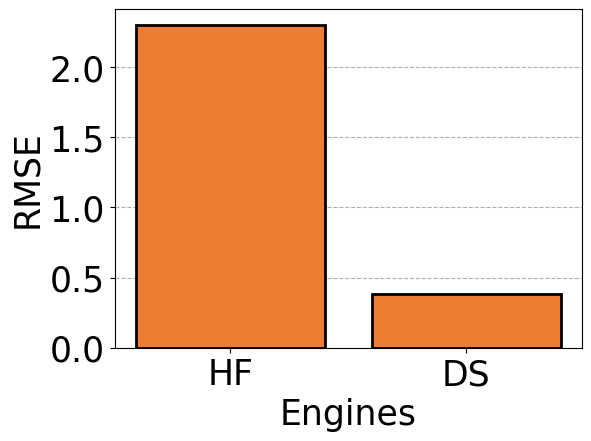}
		\caption{Error of 128 iterations.}
		\label{fig:error-128}
	\end{subfigure}
	\caption{Serving time estimation error of various inference engines, where HF and DS are short for huggingface-transformers and deepspeed-inference, respectively.}
	\vspace{-0.2cm}
	\label{fig:estimation_error}
\end{figure}

For the prefill latency $T_{prefill}(N, L_i)$, we can observe that when the batch size is fixed, $T_{prefill}(N, L_i)$ increases linearly with the batch input length $L_i$ from Fig. \ref{fig:prefill-input-length}. Besides, as shown in Fig. \ref{fig:prefill-batch-size}, when the input length is fixed, the $T_{prefill}(N, L_i)$ increases linearly with the batch size $N$ as well. For HF, we observe the same phenomenon. Therefore, we use the following function to fit $T_{prefill}(N, L_i)$.
\begin{equation}
	\label{eq:prefill-fit}
	T_{prefill}(N, L_i)=p_1 \cdot N \cdot L_i + p_2\cdot N + p_3 \cdot L_i + p_4,
\end{equation}
where $p_1$, $p_2$, $p_3$, and $p_4$ are parameters to be fitted. Under this function, $T_{prefill}(N, L_i)$ is a linear function of the batch input length $L_i$ when the batch size $N$ is fixed. Besides, when $L_i$ is fixed, $T_{prefill}(N, L_i)$ is also a linear function of $N$, which is consistent with our observations. 
We leverage the profiled prefill latency data of various engines to fit the function. Fig. \ref{fig:error-iteration} depicts the fitting error metric, root-mean-square error (RMSE), which represents the average gap between the estimated and actual latency. As shown in Fig. \ref{fig:error-iteration}, the average gap is merely 0.16s for HF and less than 0.04s for DS, achieving an accurate estimation.

For the per-iteration decoding latency $\tau_{decode}(l, N)$. we can observe from Fig. \ref{fig:decode-input-length} that when the batch size is fixed, $\tau_{decode}(l, N)$ rises with fluctuations as the context length increases. Besides, as shown in Fig. \ref{fig:decode-batch-size}, when the context length is fixed, $\tau_{decode}(l, N)$ also shows a rising trend with the increase of the batch size, and when the context length is large, the tendency tends to be linear. For HF, we also observe the same phenomenon. Hence, we try to estimate the per-iteration latency the same way we estimate the prefill latency with
\begin{equation}
	\label{eq:decode-fit}
	\tau_{decode}(l, N)=d_1 \cdot N \cdot l + d_2\cdot N + d_3 \cdot l + d_4,
\end{equation}
where $d_1$, $d_2$, $d_3$, and $d_4$ are parameters to be fitted. Fig. \ref{fig:error-iteration} presents the estimation error of $\tau_{decode}(l, N)$ for various inference engines. We can find that the fitting error is negligible, which confirms that the estimation is accurate.

By incorporating Eq. (\ref{eq:decode-fit}) into Eq. (\ref{eq:decoding-time}), we can obtain the estimated serving time of the decoding phase. Adding the decoding latency to the prefill latency computed by Eq. (\ref{eq:prefill-fit}), we can get an estimate of the batch serving time. We also evaluate the estimation error of overall batch serving time for an iteration number of 128 under various batch input lengths and batch sizes. As shown in Fig. \ref{fig:error-128}, the average error is merely 2.3s for HF and less than 0.4s for DS, which confirms that the estimation is accurate.

From Fig. \ref{fig:estimation_error}, we can find that both the per-iteration error and multi-iteration error of HF are much higher than those of DS. This is because DS leverages customized CUDA kernels to accelerate the computation of transformer blocks, so its latency bases are much smaller than that of HF, and hence its average estimation error is smaller than that of HF. 
Moreover, since the prefill latency is typically much larger than the per-iteration decoding latency, the average estimation error of the prefill latency is also larger than that of the decoding latency.

SCLS sets the iteration number limit for static batching to a small slice length $S$. When the number of iterations reaches $S$, the batch serving ends regardless of whether or not all requests have generated the EOS token. Therefore, we can consider the batch generation length $L_o$ to be $S$ for batch serving. Hence, the batch serving time of SCLS can be accurately and quickly estimated by $T_{serve}(N, L_i, S)$, where $N$ and $L_i$ can be directly obtained before batch serving. 

If all requests in the batch generate the EOS token before the generation length reaches $S$, the batch serving will finish prematurely and the estimated serving time will be inaccurate. However, since $S$ is always set small, the likelihood of the premature finish happening is small, and the performance of SLCS won't be affected. We verify this in Section \ref{sec:experiment}.

The serving time estimator is trained on offline profiled data and then applied for online prediction. Compared with exhaustively profiling the serving time under various batch input lengths and batch sizes for a given slice length, such an analytical model can be established merely by profiling the per-iteration prefill and decoding latency, greatly reducing the profiling overhead. Furthermore, the online estimation is also simple and efficient.

\subsection{Practical Memory Usage Estimation}
\label{sec:solution-memory-estimation}
During LLM serving, GPU memory is mainly occupied by the parameters of the LLM and the key-value cache of requests. The memory consumed by the key-value cache is determined by the batch size, batch input length, and batch generation length. Since for static batching, both the pad tokens and generated invalid tokens produce key-value cache, the memory consumed by key-value cache is calculated by
\begin{equation}
	\label{eq:kvcache-mem}
	M_{kv}(N, L_i, L_o)=(L_i + L_o)\cdot N \cdot \Delta,
\end{equation}
where $N$, $L_i$, and $L_o$ respectively represent the batch size, the batch input length, and the batch generation length. $\Delta$ denotes the per-token memory usage of key and value tensors, which is determined by the model architecture.

To avoid OOM errors during serving, the memory usage of the key-value cache can not exceed the maximal available GPU memory $M_{ava}$, which can be calculated by 
\begin{equation}
	\label{eq:ava-mem}
	M_{ava}=M_{cap}-M_{model}-M_{engine},
\end{equation}
where $M_{cap}$ is the GPU memory capacity, $M_{model}$ is the memory consumed by LLM parameters, and $M_{engine}$ is the memory utilized by the inference engine to store engine-specific data. 

Theoretically, given the batch size $N$, the batch input length $L_i$, and the slice length $S$, in SCLS, the memory estimator can determine whether the OOM errors will occur during batch serving by simply checking if the following constraint hold
\begin{equation}
	\label{eq:mem-constraint}
	M_{kv}(N, L_i, S)\le M_{ava}.
\end{equation}

In addition, by incorporating Eq. (\ref{eq:kvcache-mem}) into Eq. (\ref{eq:mem-constraint}) and rearranging it, we can obtain Eq. (\ref{eq:max-batchsize}), which indicates that given the batch input length $L_i$ and slice length $S$, limiting the batch size to be less than $N_{max}$ can avoid OOM errors.

\begin{equation}
	\label{eq:max-batchsize}
	N_{max}=\lfloor \frac{M_{ava}}{\Delta\cdot (L_i+S)}\rfloor.
\end{equation}

In practice, there is a gap between the actual GPU memory usage and the theoretical usage for different inference engines due to the engine-specific memory management mechanism.

HF is an inference engine based entirely on pytorch \cite{paszke2019pytorch}. It produces memory fragmentation that cannot be utilized during the batch serving. Therefore, for HF, it is necessary to multiply $M_{ava}$ in Eq. (\ref{eq:mem-constraint}) by an additional coefficient $\zeta$ less than 1, and the constraint is rewritten as
\begin{equation}
	\label{eq:mem-constraint-hf}
	M_{kv}(N, L_i, S)\le \zeta\cdot M_{ava}.
\end{equation}

DS has an inflexible memory management mechanism. We experimentally find that it is unable to determine whether OOM errors will occur using Eq. (\ref{eq:mem-constraint-hf}) for DS, even if $\zeta$ is set to a small value. Therefore, for DS, for a given slice length $S$, we profile the maximal batch size for various batch input lengths and make rules for determining whether OOM will occur during batch serving.

\subsection{Serving Time-Oriented Batching}
\label{sec:solution-batching}
\begin{figure}[t]
	\centering
	\begin{subfigure}{0.33\linewidth}
		\centering
		\includegraphics[width=1\linewidth]{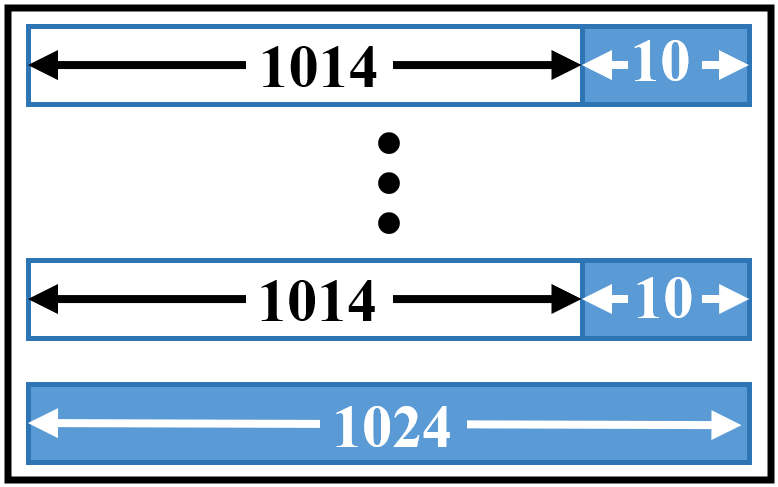}
		\caption{Together batching.}
		\label{fig:batch-together}
	\end{subfigure}
	\centering
	\begin{subfigure}{0.325\linewidth}
		\centering
		\includegraphics[width=1\linewidth]{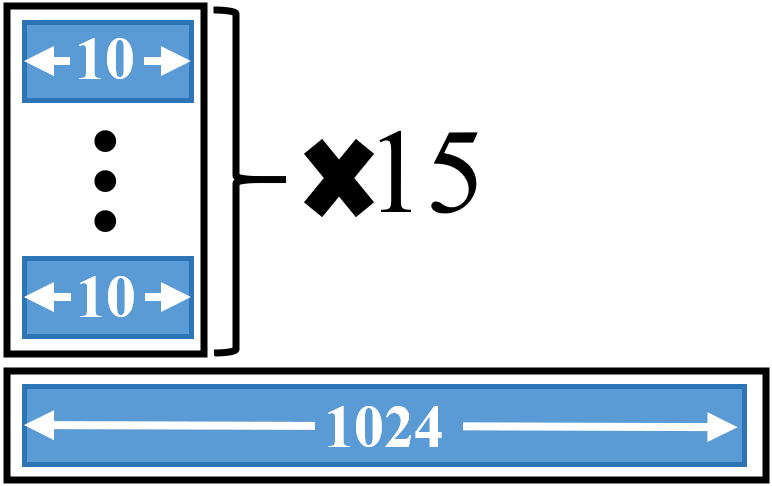}
		\caption{Separate batching.}
		\label{fig:batch-separately}
	\end{subfigure}
	\centering
	\begin{subfigure}{0.28\linewidth}
		\centering
		\includegraphics[width=1\linewidth]{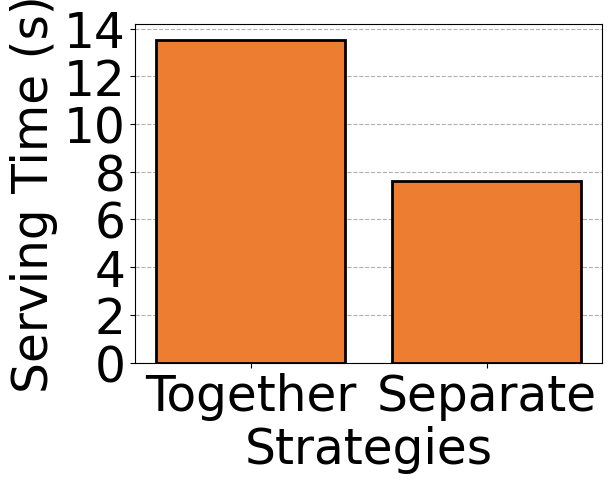}
		\caption{Serving time.}
		\label{fig:latency_diff}
	\end{subfigure}
	\caption{Batching example. Under a given slice length of 128, various batching strategies have a large difference in the serving time. In sub-figure (a), white and blue grids represent pad and input tokens, respectively.}
	\vspace{-0.5cm}
	\label{fig:batch-example}
\end{figure}

As shown in Fig. \ref{fig:batch-example}, under a slice length of 128, given 15 requests with an input length of 10 and 1 request with an input length of 1024, if they are batched together, it takes 13.5s to finish serving for an LLaMA2-13B HF-worker deployed on one NVIDIA A100 80GB GPU. If the requests with input lengths of 10 and 1024 are divided into two separate batches, the total serving time is only 7.6s. This is because all the requests are padded to 1024 tokens when they are batched together, which greatly increases the per-iteration prefill and decoding latency for the requests with an input length of 10.

Accordingly, we propose the adaptive batcher which leverages a dynamic programming-based algorithm to partition requests into batches with the goal of minimizing the total batch serving time. By aiming to reduce serving time, the algorithm can make an excellent trade-off between mitigating padding and increasing batch size.

Pseudocode of the batching algorithm is presented in Algorithm \ref{alg:batching}. It first sorts requests in ascending order by their input lengths. Then the state array $\mathcal{T}$ is created, where $\mathcal{T}[i]$ denotes the minimal total batch serving time for the first $i$ requests. The initial state $\mathcal{T}[0]=0$ indicates that the total serving time is 0 when there is no request to batch. The position array $\mathcal{P}$ stores positions for partitioning requests into batches.

During dynamic programming, the algorithm sequentially traverses each request. For each request, the algorithm first treats it as a separate batch and then tries to batch it together with preceding requests to minimize the total batch serving time while guaranteeing that OOM errors will not occur. The state transfer function of dynamic programming is 
\begin{equation}
	\label{eq:state-transfer}
	\mathcal{T}[i]=\min\limits_{0<j\le i} (\mathcal{T}[j-1] + T_{serve}(i-j+1, L_i, S)),
\end{equation}
where $i-j+1$ computes the batch size.
Since requests are sorted by the request input length at the beginning when traversing to the $i$th request, its input length is the maximal input length of the first $i$ requests, and hence its input length can be utilized as the batch input length $L_i$ to calculate the estimated batch serving time $T_{serve}$. After the dynamic programming, the requests are divided into batches according to the recorded splitting positions.

In the batching procedure, whether OOM errors will occur is determined by the memory usage estimator, and the batch serving time is estimated by the serving time estimator. 

\begin{algorithm}[t!]
	\caption{Serving Time-Oriented Batching}
	\label{alg:batching}
	\KwIn{$\mathcal{R}$: a list of requests; $S$: the slice lenth}
	\KwOut{$\mathcal{L}$: a list of batches}
	\DontPrintSemicolon
	Sort $\mathcal{R}$ by the request input length in ascending order
	
	n $\gets$ the number of requests in $\mathcal{R}$
	
	\textcolor{blue}{\tcp{States denoting total serving time}}
	
	$\mathcal{T}$ $\gets$ array of zeroes[0..n]
	
	\textcolor{blue}{\tcp{Positions for splitting batches}}
	
	$\mathcal{P}$ $\gets$ array of zeroes[0..n]
	
	\For{$i \gets 1$ \KwTo $n$}{
		$\mathcal{P}[i] \gets i - 1$ \textcolor{blue}{\tcc*[l]{Separate the $i$th request into a single batch}}
		
		$L_{i} \gets$ input length of the request $\mathcal{R}[i - 1]$
		
		$\mathcal{T}[i] \gets \mathcal{T}[i - 1]$ + $T_{serve}(1, L_i, S)$
		
		$j \gets i - 1$
		
		\While{$j > 0$ \textbf{and} \textbf{not} $OOM(i - j + 1, L_i, S)$}{
			
			$t \gets \mathcal{T}[j - 1]$ + $T_{serve}(i - j + 1, L_{i}, S)$
			
			\If{$t < \mathcal{T}[i]$}
			{ 
				$\mathcal{T}[i] \gets t$ \textcolor{blue}{\tcc*[l]{Update the total serving time}}
				
				$\mathcal{P}[i] \gets j - 1$ \textcolor{blue}{\tcc{Record the splitting position}}
			}
			
			$j \gets j - 1$\;
		}
	}
	$\mathcal{L} \gets$ empty list \textcolor{blue}{\tcp*[l]{Now, $i$ is equal to n}}
	\While{$i > 0$}{
		$p \gets \mathcal{P}[i]$
		
		Batch $\mathcal{R}[p:i]$ together and append the batch to $\mathcal{L}$
		
		$i \gets p$
	}
	\Return{$\mathcal{L}$}
\end{algorithm}

\subsection{Balanced Load-Oriented Offloading}
\label{sec:solution-offloading}
The offloader achieves load balancing among multiple workers using a max-min-based algorithm. The load of a worker is defined as the time it takes to serve all the batches in its local queue, and the initial load of each worker is 0.

The offloader offloads batches to workers one by one. At each offload, it schedules the batch with the longest estimated serving time to the worker with minimal load and then updates the worker's load with 
\begin{equation}
	\label{eq:load-update}
	T_{load}(w)\gets T_{load}(w) + T_{serve}(N, L_i, S),
\end{equation}
where $T_{load}(w)$ represents the load of the least loaded worker $w$ and $T_{serve}(N, L_i, S)$ denotes the estimated serving time of the batch to be offloaded, where $N$, $L_i$, and $S$ are the batch size, batch input length, and slice length, respectively.

After each time a worker completes batch serving, SCLS subtracts the estimated serving time of the batch it just served from its load. This can prevent the estimation error of serving time from accumulating in the worker's load.

\subsection{Adaptive Schedule Interval Update}
\label{sec:solution-interval-update}
SCLS periodically fetches requests from the request pool and hands them over to the adaptive batcher at a time interval of $T$. After each batch offloading, SCLS updates $T$ by
\begin{equation}
	\label{eq:interval-update}
	T\gets \max(\lambda \cdot \min\limits_w (T_{load}(w)), \Gamma),
\end{equation}
where $\Gamma$ is a pre-defined minimal time interval, $T_{load}(w)$ is the load of worker $w$, and $\lambda$ is a factor less than 1.

When the load of each worker is light, $T$ decreases so that requests can be processed by workers in time without waiting for a long time in the request pool, thus reducing the request response time. When there are a lot of batches waiting in the local queue of each worker, $T$ increases so that the adaptive batcher is highly likely to batch more requests together, thus increasing the throughput. Due to the serving time estimation error,
$\lambda$ is introduced to prevent workers from being idle when the load is over-estimated, and $\Gamma$ is introduced to avoid feeding too few requests to the adaptive batcher when the load is under-estimated.

\begin{algorithm}[t]
	\caption{DeepSpeed-Inference OOM Judgment}
	\label{alg:deepspeed-oom}
	\KwIn{ $L_i$:batch input length; 
		$S$: slice length; $N$: batch size}
	\KwOut{An expression that OOM will occur if true}
	\DontPrintSemicolon
	
	$L$ $\gets$ $L_i$ + $S$ \textcolor{blue}{\tcc{$L\le 2048$ under the experimental settings}}
	
	\uIf{L $>$ 1024}{
		
		\Return{$N>12$}
		
	}
	\uElseIf{L $>$ 512}{
		
		\Return{$N>22$}
		
	}
	\Else{
		
		\Return{$N>28$} \textcolor{blue}{\tcc{Total token number $L\le 512$}}
		
	}
\end{algorithm}

\vspace{-0.4cm}
\section{Experimental Evaluation}
\label{sec:experiment}
\vspace{-0.15cm}
\subsection{Experiment Setup}
\vspace{-0.1cm}
Experiments are conducted in the production environment at Ant Group, where 8 LLaMA-13B LLM instances are deployed with 8 NVIDIA A100 80GB GPUs connected over NVLink. We use LLaMA2-13B in the experiments because LLaMA's model architecture is representative and has been adopted by many LLMs. Besides, 13B-parameter LLMs can meet the needs of most real-world application scenarios in the production environment. We take the request throughput, average and tail response time as metrics to evaluate the performance of SCLS, following existing work on LLM serving \cite{yu2022orca, kwon2023efficient, wu2024dlora}.

We implement SCLS, DS-worker and HF-worker with Python 3.8. For the serving time estimator, we use \verb|curve_fit| function from the scipy library \cite{virtanen2020fundamental} to fit Eq. (\ref{eq:prefill-fit}) and Eq. (\ref{eq:decode-fit}). For the memory usage estimator, $\zeta$ in Eq. (\ref{eq:mem-constraint-hf}) is set to 0.9 for HF. For DS, SCLS determines if OOM errors will occur by rules that are presented in Algorithm \ref{alg:deepspeed-oom}. To fetch more requests while keeping workers busy, SCLS dynamically adjusts the schedule time interval $T$ by Eq. (\ref{eq:interval-update}), where $\lambda$ is set to 0.5, $\Gamma$ is set to 6s and 3s for HF and DS according to their inference speed, respectively. Moreover, the slice length is set to 128 if not specified. 

Requests from CodeFuse's traces are sent in the order they actually arrived for 10 minutes. Arrival times are generated using Poisson distribution with various request rates. We set both the maximal raw request input length and the maximal generation length limit to 1024, covering length range settings of existing studies \cite{zheng2024response, jin2024s, yu2022orca, kwon2023efficient, sheng2024fairness}.

We compare SCLS with SLS and ILS schedulers. Since HF only supports static batching, for HF, we only compare SCLS with SLS. For DS, we compare SCLS with SLS and Deepspeed-Fastgen, an ILS scheduler implemented on top of DS. We don't compare SCLS with vLLM \cite{vllm} because vLLM does not support static batching, and hence SCLS can not achieve the same kernel-level optimization with vLLM, which makes it challenging to achieve a fair comparison from a pure scheduling perspective. Besides, Deepspeed-Fastgen has been shown to outperform vLLM's performance \cite{deepspeed-fastgen}. Therefore, we think it reasonable and convincing to compare SCLS with Deepspeed-Fastgen. 
\begin{itemize}
	\item \textbf{Sequence-level Scheduling (SLS)} \cite{tensorflowserving, triton}: The scheduler offloads requests to workers using the round-robin policy. Workers serve received requests in an FCFS manner with a fixed batch size. Since different engines have various $M_{engine}$, for HF and DS workers, the fixed batch size is set to 16 and 12 to avoid OOM errors, respectively.
	\item \textbf{Iteration-level Scheduling (ILS)} \cite{deepspeed-fastgen}: Deepspeed-FastGen, the ILS scheduler built on top of DS, achieves a low latency with continuous batching \cite{yu2022orca}, paged attention\cite{kwon2023efficient}, and chunked-prefill \cite{agrawal2024taming} techniques. However, it utilizes the round-robin policy to offload requests and adopts a conservative memory management mechanism that limits the number of parallel-processing requests. 
	Since Deepspeed-FastGen is also implemented based on DS, when SCLS is integrated with DS, SCLS and ILS have the same kernel-level optimization, so a fair comparison can be made from a pure scheduling perspective.
\end{itemize}

\begin{figure*}[t]
	\setlength{\belowcaptionskip}{-0.15cm} 
	\centering
	\begin{subfigure}{0.32\linewidth}
		\centering
		\includegraphics[width=1\linewidth]{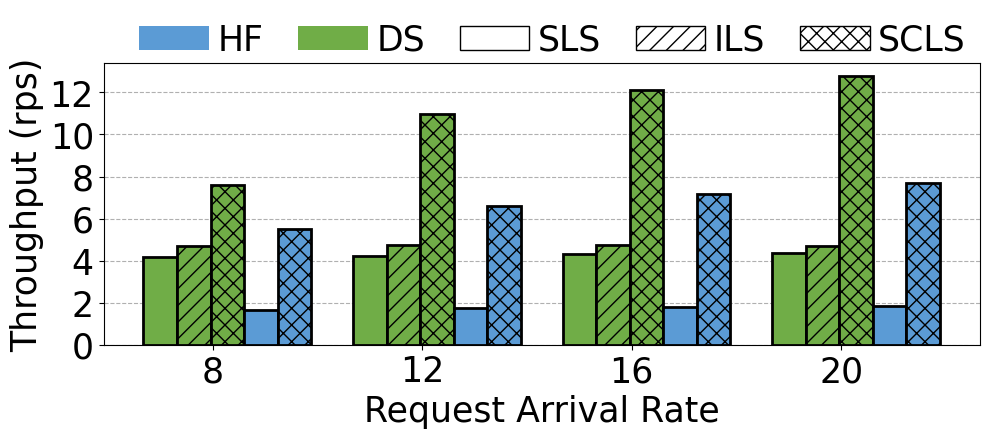}
		\caption{Request throughput.}
		\label{fig:eval_request_tp}
	\end{subfigure}
	\centering
	\begin{subfigure}{0.33\linewidth}
		\centering
		\includegraphics[width=1\linewidth]{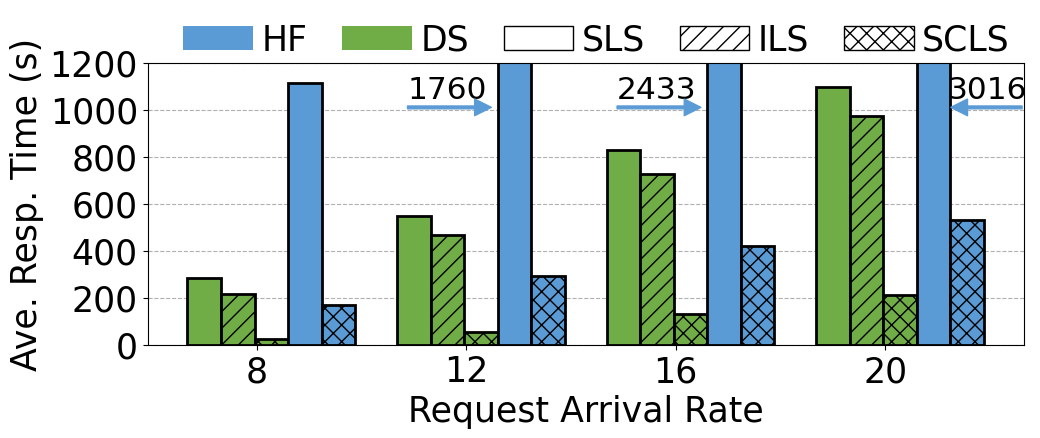}
		\caption{Average response time.}
		\label{fig:eval_avg_response}
	\end{subfigure}
	\centering
	\begin{subfigure}{0.33\linewidth}
		\centering
		\includegraphics[width=1\linewidth]{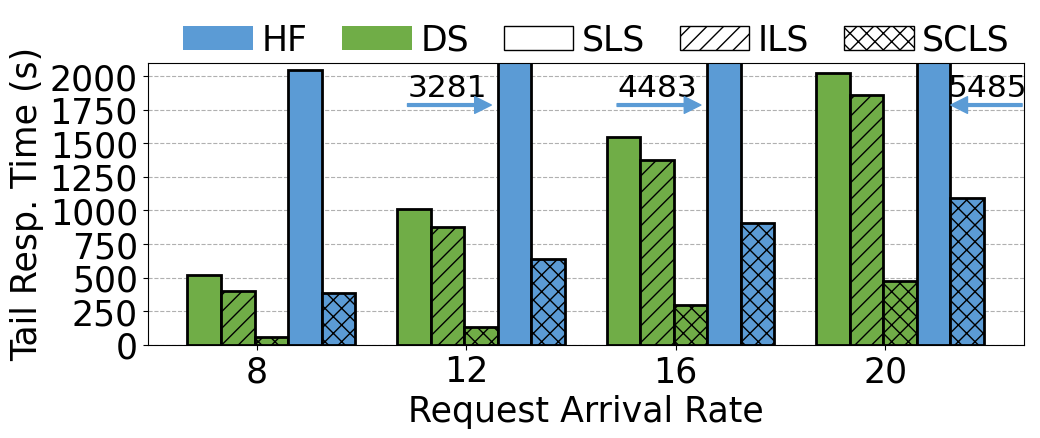}
		\caption{Tail (95\%) response time.}
		\label{fig:eval_tail_response}
	\end{subfigure}
	\caption{Request throughput, average response time and tail response time under various arrival rates.}
	\label{fig:overall_perfprmance}
	\vspace{-0.25cm}
\end{figure*}

\begin{figure*}[t]
	\setlength{\belowcaptionskip}{-0.15cm} 
	\centering
	\begin{subfigure}{0.33\linewidth}
		\centering
		\includegraphics[width=1\linewidth]{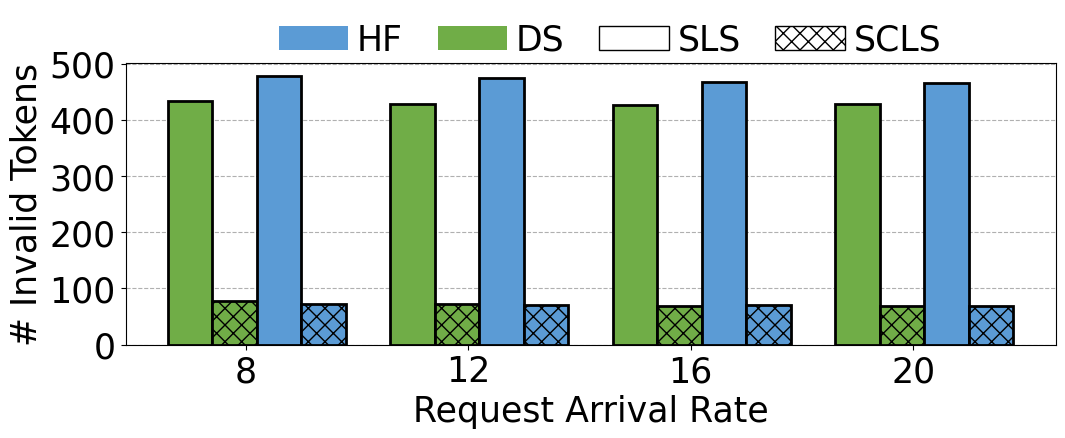}
		\caption{Average number of invalid tokens.}
		\label{fig:eval_invalid_tokens}
	\end{subfigure}
	\centering
	\begin{subfigure}{0.325\linewidth}
		\centering
		\includegraphics[width=1\linewidth]{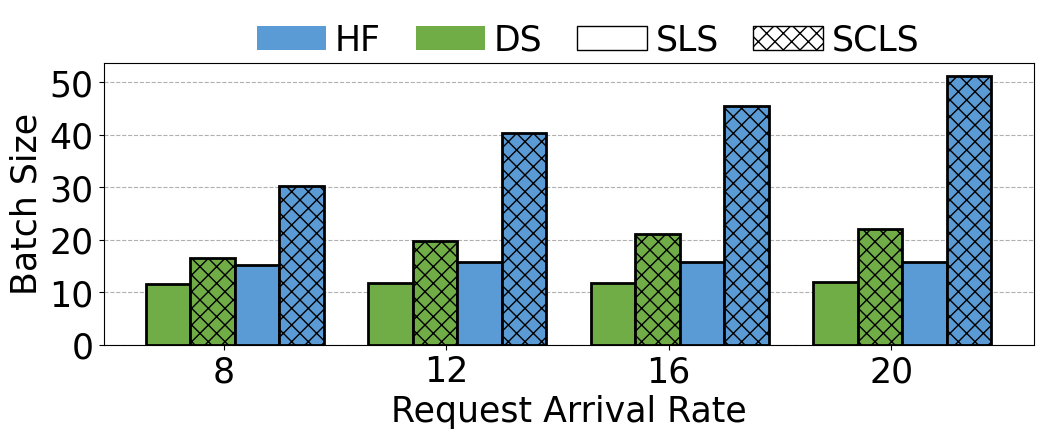}
		\caption{Average batch size.}
		\label{fig:eval_batch_size}
	\end{subfigure}
	\centering
	\begin{subfigure}{0.33\linewidth}
		\centering
		\includegraphics[width=1\linewidth]{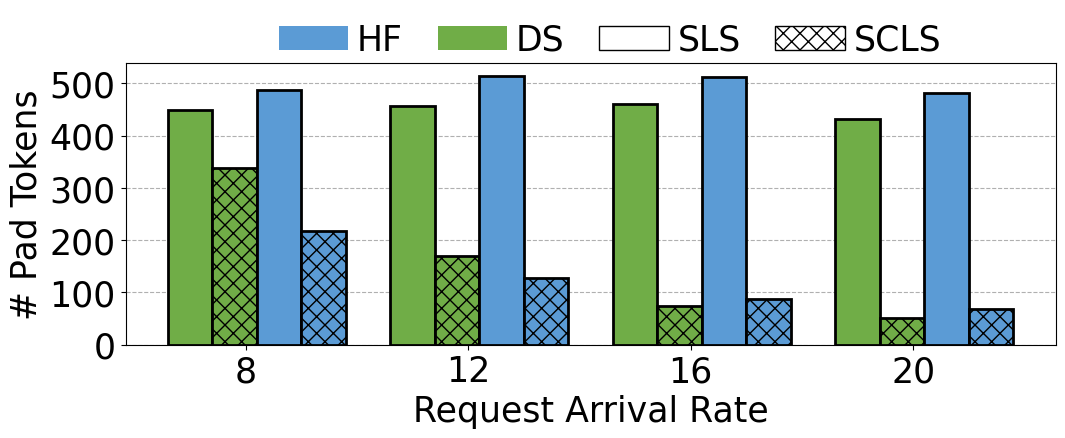}
		\caption{Average number of pad tokens.}
		\label{fig:eval_pad_tokens}
	\end{subfigure}
	\caption{Dive into the superiority of SCLS.}
	\label{fig:superiority}
	\vspace{-0.25cm}
\end{figure*}

\begin{figure}[t]
	\setlength{\belowcaptionskip}{-0.15cm} 
	\centering
	\begin{subfigure}{0.465\linewidth}
		\centering
		\includegraphics[width=1\linewidth]{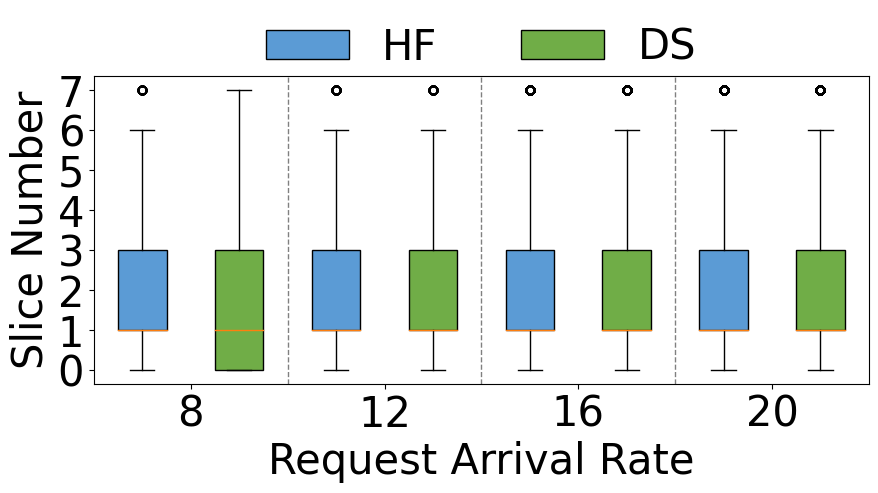}
		\caption{Slice number distribution.}
		\label{fig:eval_slice_number}
	\end{subfigure}
	\centering
	\begin{subfigure}{0.5225\linewidth}
		\centering
		\includegraphics[width=1\linewidth]{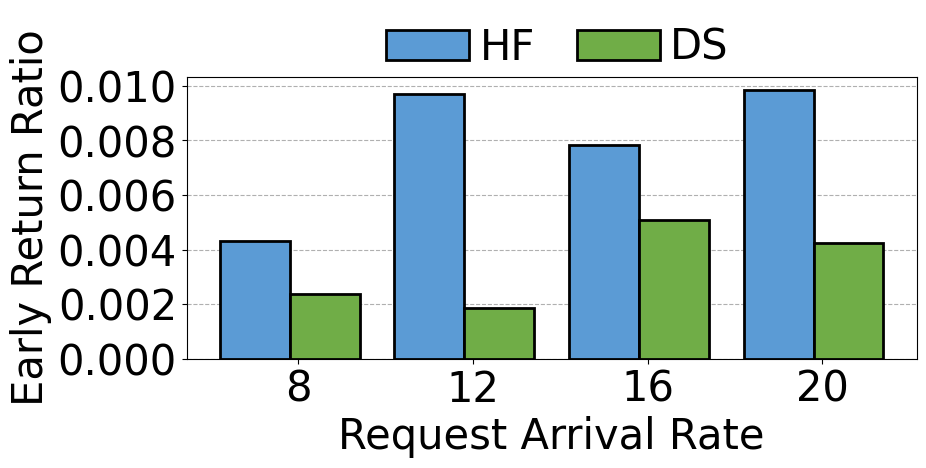}
		\caption{Early return ratio.}
		\label{fig:eval_early_return_ratio}
	\end{subfigure}
	\caption{Dive into the overhead of SCLS.}
	\label{fig:overahead}
	\vspace{-0.25cm}
\end{figure}

\vspace{-0.33cm}
\subsection{Overall Performance}
\vspace{-0.12cm}
We compare the performance of SCLS to the baseline schedulers in terms of request throughput and tail response time. The experimental results are depicted in Fig. \ref{fig:overall_perfprmance}, where different colors represent different inference engines, and various hatch styles represent various scheduling techniques. We can see that SCLS always outperforms SLS and ILS when integrating with various inference engines under different request arrival rates, as shown in Fig. \ref{fig:overall_perfprmance}.

For HF, SCLS can increase the throughput by 232.3\% to 315.8\%,  decrease the average response time by 82.4\% to 84.9\%, and decrease the tail response time by 79.8\% to 81.0\% compared with SLS. For DS, SCLS increases the throughput by 82.5\% to 191.9\%, reduces the average response time by 80.8\% to 91.1\%, and decreases the tail response time by 76.6\% to 88.9\% compared with SLS.

Compared with ILS, SCLS can support larger batch sizes without causing OOM errors, whereas Deepspeed-FastGen adopts a conservative memory management policy by limiting the number of parallel-processing requests. As a result, SCLS can improve the throughput by 61.6\% to 171.0\%, reduce the average request response time by 78.4\% to 88.3\%, and reduce the tail response time by 74.6\% to 85.8\% compared with ILS when integrated with DS.

Compared with SLS, the superiority of SCLS mainly comes from three sources. The first is to return completed requests in time, thus avoiding request waiting and greatly reducing the number of generated invalid tokens, as shown in Fig. \ref{fig:eval_invalid_tokens}.
The second is to serve requests with a large batch size to achieve high throughput, as shown in Fig. \ref{fig:eval_batch_size}. The third is to reduce the serving latency by mitigating the padding overhead using the adaptive batcher as shown in \ref{fig:eval_pad_tokens}. Due to the different memory management mechanisms of HF and DS, the maximal batch size that can be supported under the same batch input length and the same slice length is different, and therefore, SCLS obtains different performance gains when integrated with various inference engines.

\subsection{SCLS Overhead}
The overhead of SCLS comes from the re-computation of the prefill phase with each reschedule. We illustrate the distribution of request reschedule number (a.k.a. slice number) under different request rates in Fig. \ref{fig:eval_slice_number}. We find that the vast majority of requests have less than three reschedules. This is because their request generation lengths are small and can complete generation within three slices.
Although the total serving time of requests with long generation lengths is increased due to multiple re-computations of the prefill phase, they also benefit from SCLS as their queuing time is greatly reduced due to increased throughput.

In Section \ref{sec:solution-time-estimation}, we mentioned that if all requests generate EOS before the generation length reaches the slice length, the requests will be returned early, which leads to inaccurate time estimation. To verify that this happens very rarely, in Fig. \ref{fig:eval_early_return_ratio}, we calculate the proportion of these early returned batches to the total number of served batches. 
We can find that early return happens rarely, with a maximal proportion less than 1\%. Hence, the early return has a negligible effect on the performance of SCLS.

\vspace{-0.4cm}
\subsection{Load Balance}
\begin{figure}[t]
	\setlength{\belowcaptionskip}{-0.05cm} 
	\centering
	\includegraphics[width=0.9\linewidth]{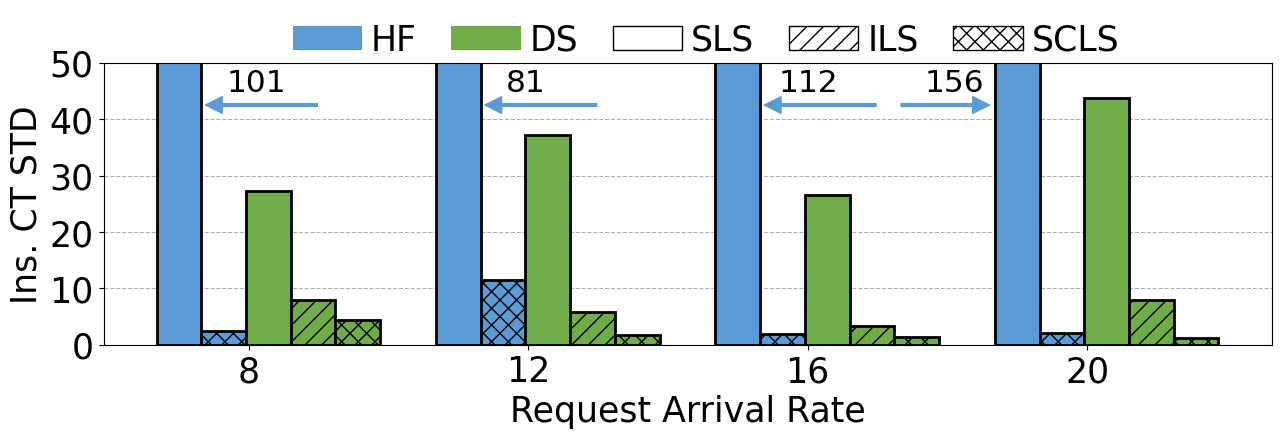}
	\caption{Load imbalance under various request rates}
	\label{fig:load_balance}
	\vspace{-0.25cm}
\end{figure}

\vspace{-0.1cm}
We calculated the standard deviation (STD) of each instance's completion time (CT) at the end of the experiment in Fig. \ref{fig:load_balance}. 
We find that the CT STD of SCLS is always the smallest, which indicates that the completion time of each instance is similar and SCLS achieves load balancing across multiple LLM instances. This is because the offloading algorithm can accurately estimate the load of workers and leverage the max-min algorithm to reduce the load gap between workers. 

Without knowing the request generation length, SLS and ILS leverage the round-robin policy to offload requests, which may schedule requests with long generation lengths and requests with short generation lengths to different instances in a period of time. Such an unbalanced workload accumulates over time and leads to different instance completion times, Therefore, they have a higher standard deviation of instance completion time than SCLS.

\begin{figure*}[t]
	\setlength{\belowcaptionskip}{-0.15cm} 
	\centering
	\begin{subfigure}{0.32\linewidth}
		\centering
		\includegraphics[width=1\linewidth]{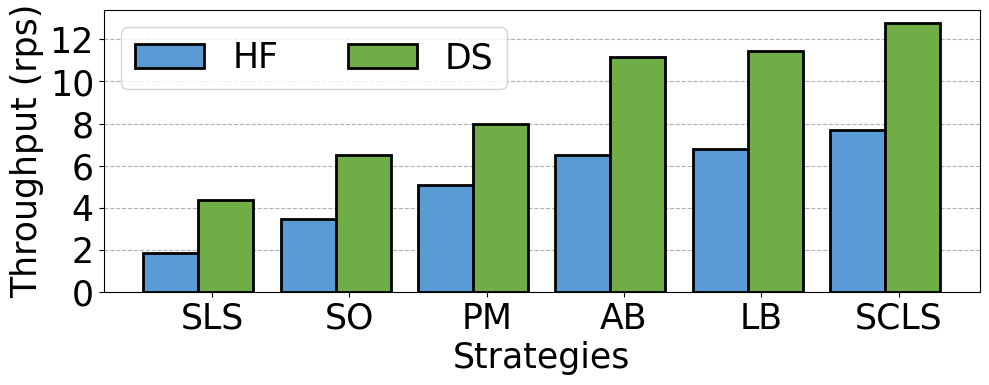}
		\caption{Request throughput.}
		\label{fig:ablation_request_tp}
	\end{subfigure}
	\centering
	\begin{subfigure}{0.33\linewidth}
		\centering
		\includegraphics[width=1\linewidth]{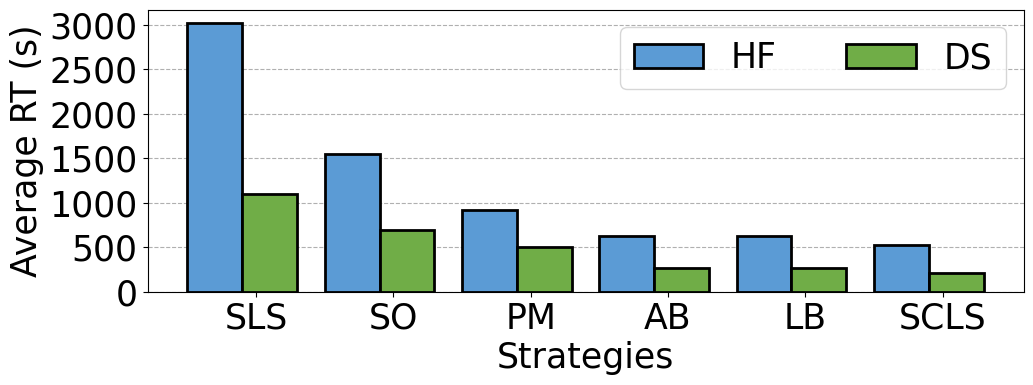}
		\caption{Average response time (RT).}
		\label{fig:ablation_avg_response}
	\end{subfigure}
	\centering
	\begin{subfigure}{0.33\linewidth}
		\centering
		\includegraphics[width=1\linewidth]{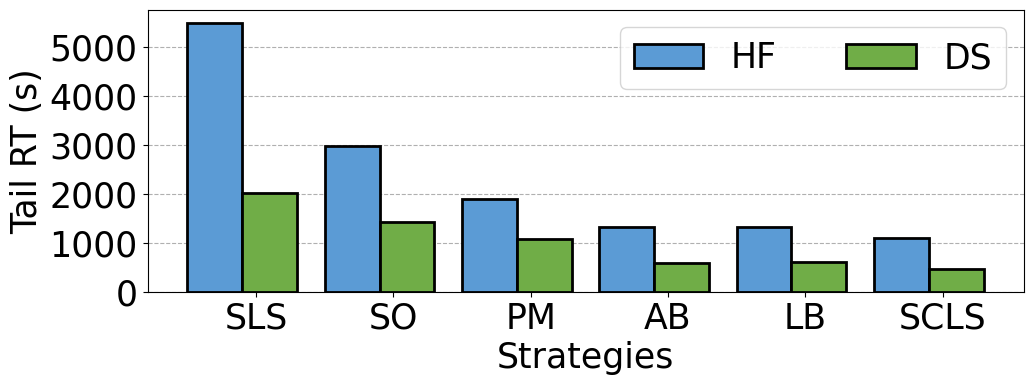}
		\caption{Tail (95\%) response time (RT).}
		\label{fig:ablation_tail_response}
	\end{subfigure}
	\caption{Request throughput, average response time and tail response time under various strategies.}
	\label{fig:ablation_study}
	\vspace{-0.25cm}
\end{figure*}

\begin{figure*}[t]
	\setlength{\belowcaptionskip}{-0.05cm}
	\centering
	\begin{subfigure}{0.33\linewidth}
		\centering
		\includegraphics[width=1\linewidth]{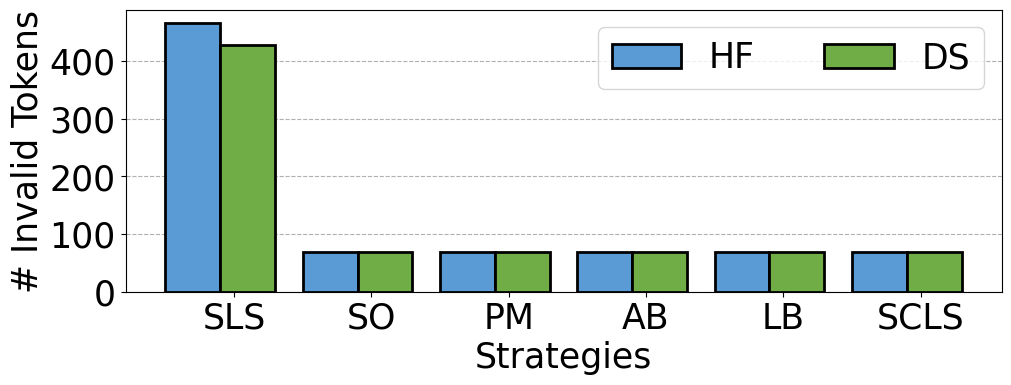}
		\caption{Average number of invalid tokens.}
		\label{fig:ablation_invalid_tokens}
	\end{subfigure}
	\centering
	\begin{subfigure}{0.325\linewidth}
		\centering
		\includegraphics[width=1\linewidth]{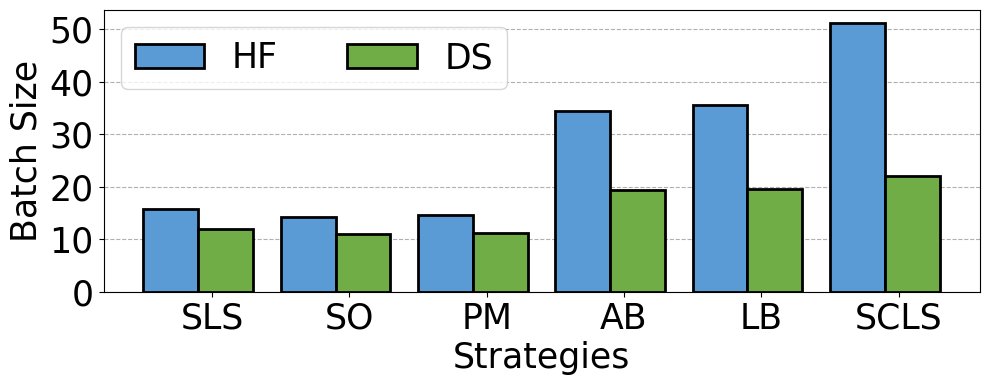}
		\caption{Average batch size.}
		\label{fig:ablation_batch_size}
	\end{subfigure}
	\centering
	\begin{subfigure}{0.33\linewidth}
		\centering
		\includegraphics[width=1\linewidth]{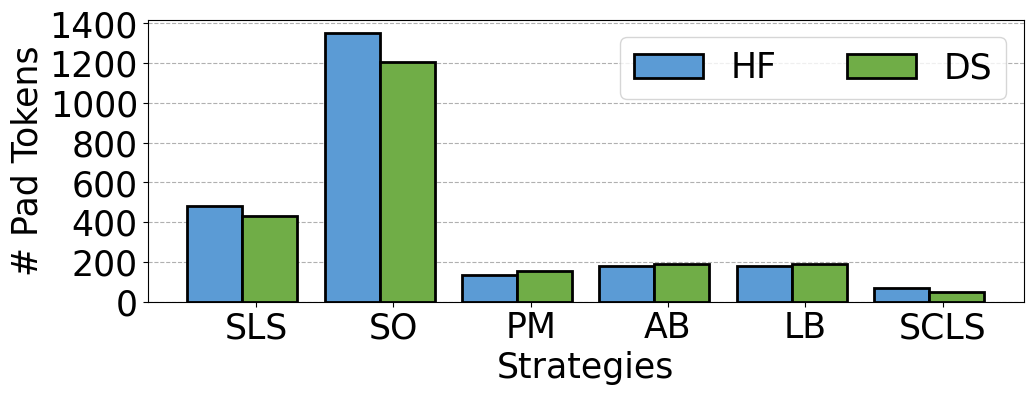}
		\caption{Average number of pad tokens.}
		\label{fig:ablation_pad_tokens}
	\end{subfigure}
	\caption{Dive into the results of ablation experiments.}
	\label{fig:ablation_contribution}
	\vspace{-0.25cm}
\end{figure*}

\vspace{-0.35cm}
\subsection{Ablation Studies}
\vspace{-0.15cm}
We demonstrate the performance gain of the design features of SCLS. We keep adding features to SLS to constitute multiple strategies and then verify their performance in terms of request throughput and response time at a request arrival rate of 20. The experimental results are presented in Fig. \ref{fig:ablation_study}.

\textbf{Slice-Only (SO)} limits the number of iterations for static batching to the slice length, returning completed requests in time, and re-sent unfinished requests to workers. It offloads requests to workers using the round-robin policy. Workers serve received requests in an FCFS manner with a fixed batch size. 
\textbf{Padding-Mitigating (PM)} implements an incomplete batching algorithm on top of SO, where the batch size is limited to 12 and 16 for DS and HF, respectively. Besides, PM fetches requests from the request pool at the fixed time interval $\Gamma$ and offloads batches to workers with the round-robin policy as well. \textbf{Adaptive-Batching (AB)} lifts the fixed batch size limitation on the basis of PM and implements the complete serving time-oriented batching algorithm. 
\textbf{Load-Balancing (LB)} adds the max-min offloading policy to AB to achieve a balanced workload across multiple LLM instances. 
By integrating the adaptive batch schedule update mechanism on LB, SCLS is constructed.

As shown in Fig. \ref{fig:ablation_study}, the performance of strategies gradually enhances as more features are integrated, where the throughput increases and the response time decreases. Compared with SLS, SO is able to return completed requests timely, thus gaining a performance boost. Compared with SO, PM continues to improve performance by reducing the additional computational overhead caused by pad tokens using the incomplete batching algorithm. Compared with PM, AB increases the batch size to improve throughput as shown in Fig.\ref{fig:ablation_batch_size}. LB achieves a balanced load across workers, thus preventing workers from being idle and fully utilizing workers' computing capability to further improve the throughput. The adaptive schedule interval update mechanism allows more requests to be fetched when the workload is heavy, giving the batching algorithm more chances to increase the batch size and reduce pad tokens.

We also present the average invalid token number of requests, the average batch size, and the average pad token number of requests for these strategies in Fig. \ref{fig:ablation_contribution}. We can see that generation slicing can significantly reduce invalid tokens, and the batching algorithm can greatly reduce pad tokens and increase the batch size. Although the number of pad tokens of AB and LB is slightly higher than that of PM, the performance of AB and LB is higher than that of PM due to the great increase in batch size, which confirms that by aiming to reduce serving time, the batching algorithm can make an excellent trade-off between mitigating padding and increasing batch size.

\begin{figure*}[b]
	\setlength{\belowcaptionskip}{-0.05cm} 
	\centering
	\begin{subfigure}{0.32\linewidth}
		\centering
		\includegraphics[width=1\linewidth]{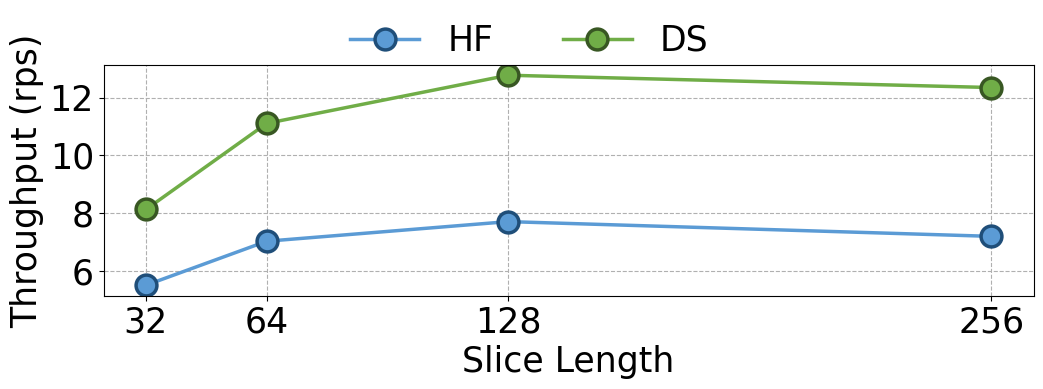}
		\caption{Request throughput.}
		\label{fig:param_request_tp}
	\end{subfigure}
	\centering
	\begin{subfigure}{0.33\linewidth}
		\centering
		\includegraphics[width=1\linewidth]{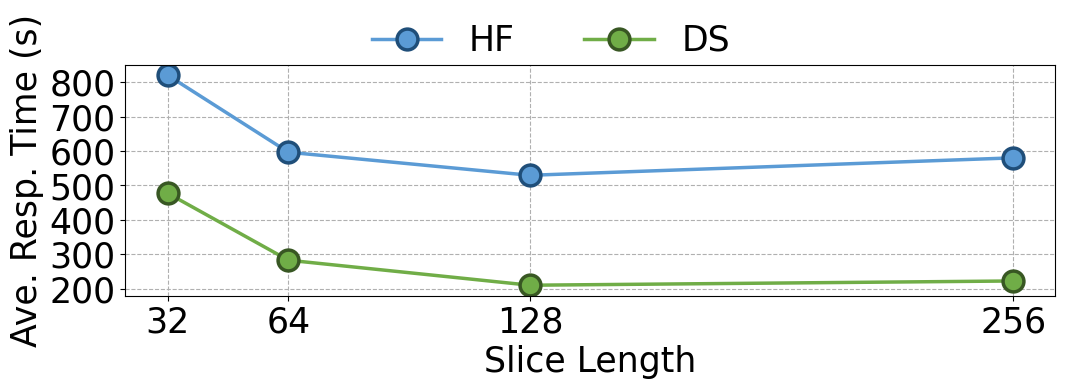}
		\caption{Average response time.}
		\label{fig:param_avg_response}
	\end{subfigure}
	\centering
	\begin{subfigure}{0.33\linewidth}
		\centering
		\includegraphics[width=1\linewidth]{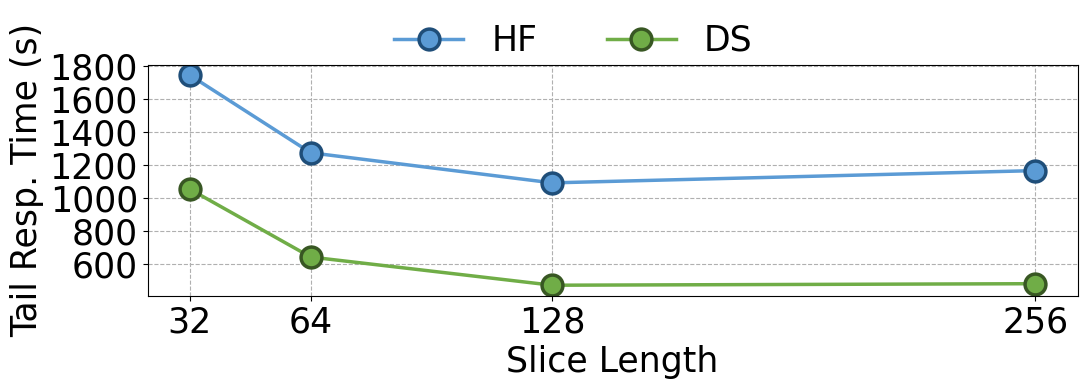}
		\caption{Tail (95\%) response time.}
		\label{fig:param_tail_response}
	\end{subfigure}
	\caption{Request throughput, average response time and tail response time under various slice lengths.}
	\label{fig:parameter_perfprmance}
	\vspace{-0.5cm}
\end{figure*}

\subsection{Impact of Slice Length}
\begin{figure*}[t]
	\setlength{\belowcaptionskip}{-0.15cm}
	\centering
	\begin{subfigure}{0.33\linewidth}
		\centering
		\includegraphics[width=1\linewidth]{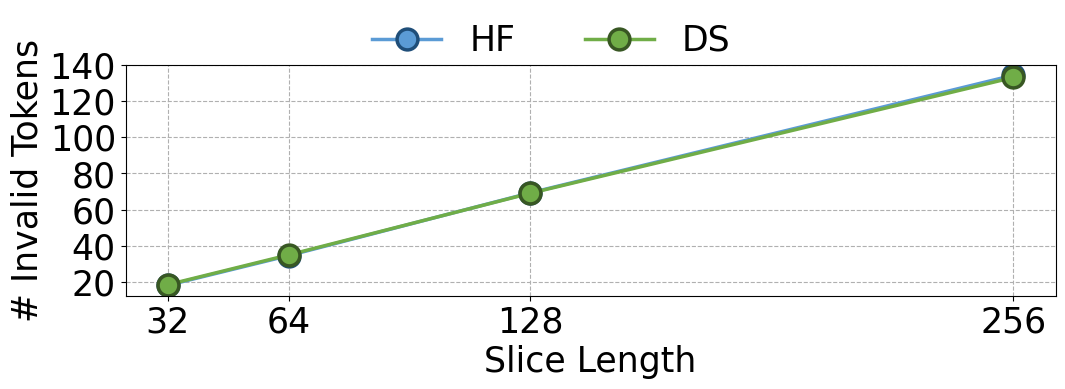}
		\caption{Average number of invalid tokens.}
		\label{fig:param_invalid_tokens}
	\end{subfigure}
	\centering
	\begin{subfigure}{0.325\linewidth}
		\centering
		\includegraphics[width=1\linewidth]{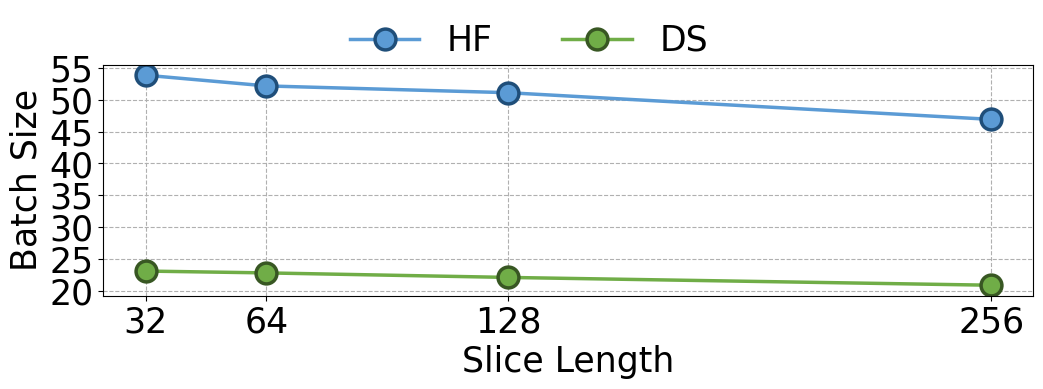}
		\caption{Average batch size.}
		\label{fig:param_batch_size}
	\end{subfigure}
	\centering
	\begin{subfigure}{0.33\linewidth}
		\centering
		\includegraphics[width=1\linewidth]{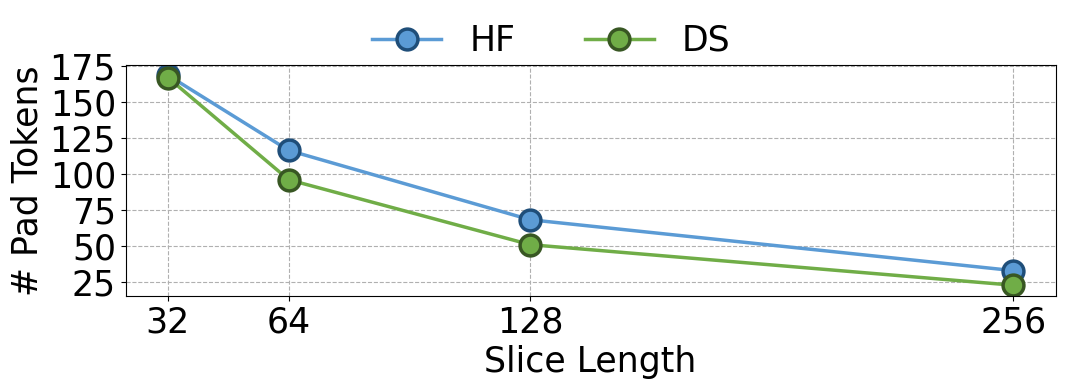}
		\caption{Average number of pad tokens.}
		\label{fig:param_pad_tokens}
	\end{subfigure}
	\caption{Dive into the slice length impact on SCLS performance .}
	\label{fig:param_dive}
	\vspace{-0.25cm}
\end{figure*}

\begin{figure}[t]
	\setlength{\belowcaptionskip}{-0.25cm} 
	\centering
	\begin{subfigure}{0.48\linewidth}
		\centering
		\includegraphics[width=1\linewidth]{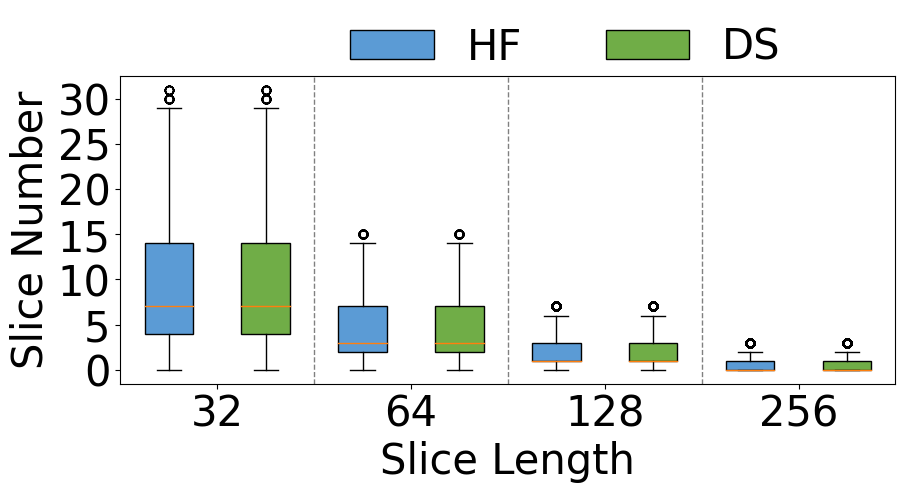}
		\caption{Slice number distribution.}
		\label{fig:param_slice_number}
	\end{subfigure}
	\centering
	\begin{subfigure}{0.5\linewidth}
		\centering
		\includegraphics[width=1\linewidth]{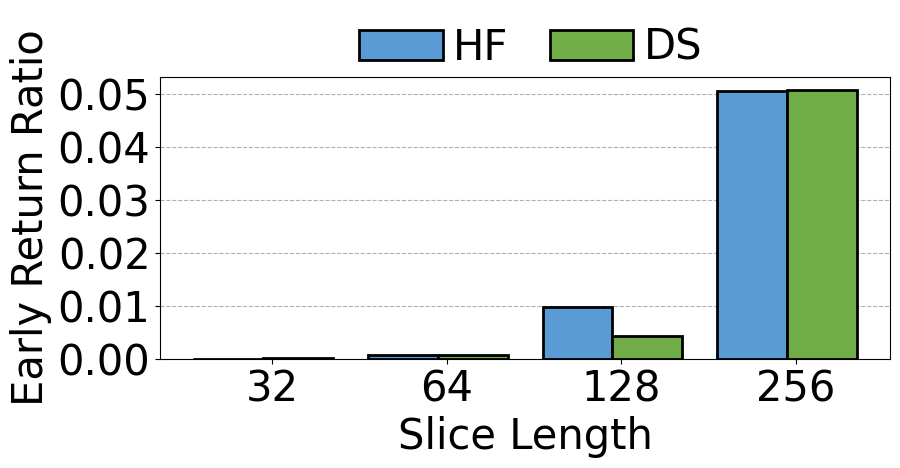}
		\caption{Early return ratio.}
		\label{fig:param_early_return_ratio}
	\end{subfigure}
	\caption{Dive into the slice length impact on SCLS overhead.}
	\label{fig:param_overahead}
	\vspace{-0.25cm}
\end{figure}

In this section, we explore the impact of slice length on the performance of SCLS. We evaluate the performance of SCLS at a request rate of 20 under various slice lengths. The experimental results are shown in Fig. \ref{fig:parameter_perfprmance}. Besides, we also present the average invalid token number of requests, the average batch size, and the average pad token number of requests in Fig. \ref{fig:param_dive}. Mover, we also depict the slice number distribution and early return ratio in Fig. \ref{fig:param_overahead} to analyze how the overhead varies with the slice length.

\begin{figure}[t]
	\setlength{\belowcaptionskip}{-0.25cm} 
	\centering
	\includegraphics[width=0.8\linewidth]{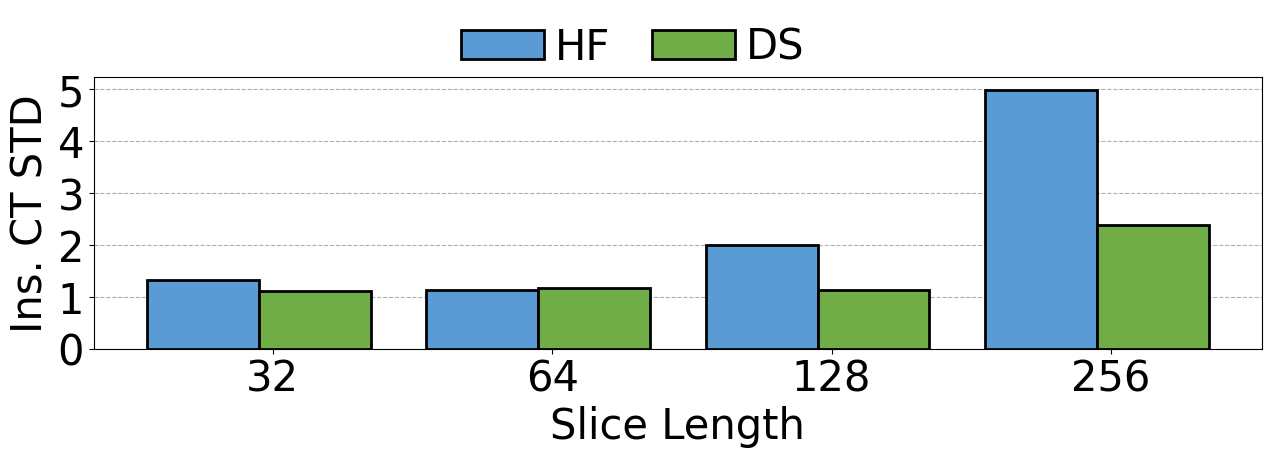}
	\caption{Load imbalance under various slice lengths}
	\label{fig:param_balance}
	\vspace{-0.25cm}
\end{figure}

From Fig. \ref{fig:parameter_perfprmance}, we can find that with the increase in slice length, the performance first increases and then decreases. This is because when the slice length is set too small, there are a lot of reschedules of requests. The reschedules cause requests to be padded multiple times, and hence the computational overhead caused by padding is severe when the slice length is set small, as shown in Fig. \ref{fig:param_pad_tokens}. Besides, the frequent reschedules cause a large number of prefill re-computations. Although with a small slice length, completed requests can be returned in a timely manner, and the batch size can be relatively large, it is not enough to make up for the computational overhead of the large amount of re-padding and prefill re-computations. When the slice length is increased, although the batch size decreases slightly, the number of reschedules decreases dramatically as shown in Fig. \ref{fig:param_slice_number}, and hence the performance is improved. However, setting the slice length too long not only further decreases the batch size as shown in Fig. \ref{fig:param_batch_size}, but also causes more serious request waiting, and more invalid tokens are generated as shown in Fig.\ref{fig:param_invalid_tokens}, so the performance decreases. Furthermore, a long slice length causes a large early return ratio as shown in Fig. \ref{fig:param_early_return_ratio}, which makes the serving time estimation inaccurate, and hence the workload becomes unbalanced as shown in Fig. \ref{fig:param_balance}. Therefore, when setting the slice length, we need to do a trade-off between the request padding, the batch size, the prefill re-computations, and the request waiting.

Since the performance of SCLS monotonically increases first and then monotonically decreases with the increase of the slice length, for application scenarios with specific request input and output distribution, we can simply use the binary search to quickly find the optimal setting of slice length with a time complexity of $\log(n)$, where $n$ is the possible maximal slice length, making the search for the optimal slice length simple and efficient.

\subsection{Handling Long-Generation Requests}
We increase the proportion of requests with generation length greater than 512 to 60\% and re-conduct the experiments at a request rate of 8. Since ILS has been demonstrated to be consistently superior to SLS, we only compare SCLS with ILS using DS as the inference engine. The results show that SCLS can improve the request throughput by 117\% compared with the ILS scheduler. Besides, the average response time is reduced by 15.7\% when the slice length increases from 128 to 256 due to greatly decreased re-prefill, which confirms that SCLS can effectively handle long generation scenarios by properly increasing the slice length.

\begin{figure}[b]
	\setlength{\belowcaptionskip}{-0.25cm} 
	\centering
	\begin{subfigure}{0.475\linewidth}
		\centering
		\includegraphics[width=1\linewidth]{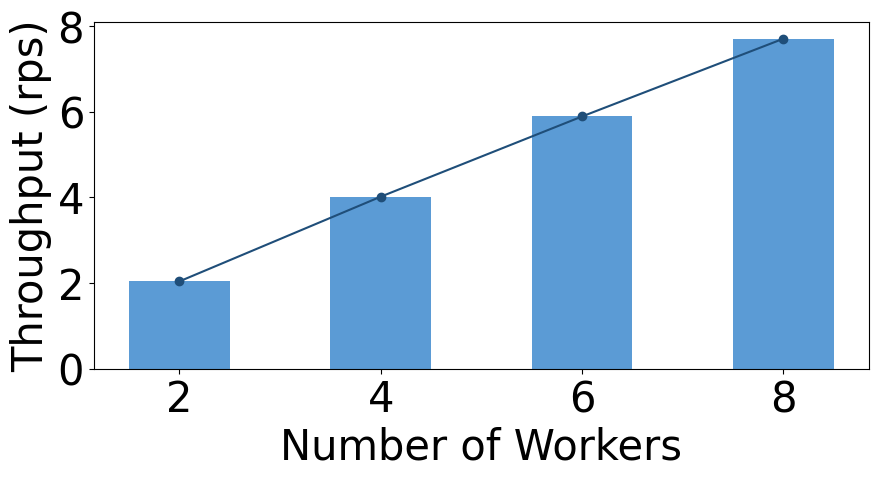}
		\caption{Huggingface-Transformers.}
		\label{fig:scale_hf}
	\end{subfigure}
	\centering
	\begin{subfigure}{0.49\linewidth}
		\centering
		\includegraphics[width=1\linewidth]{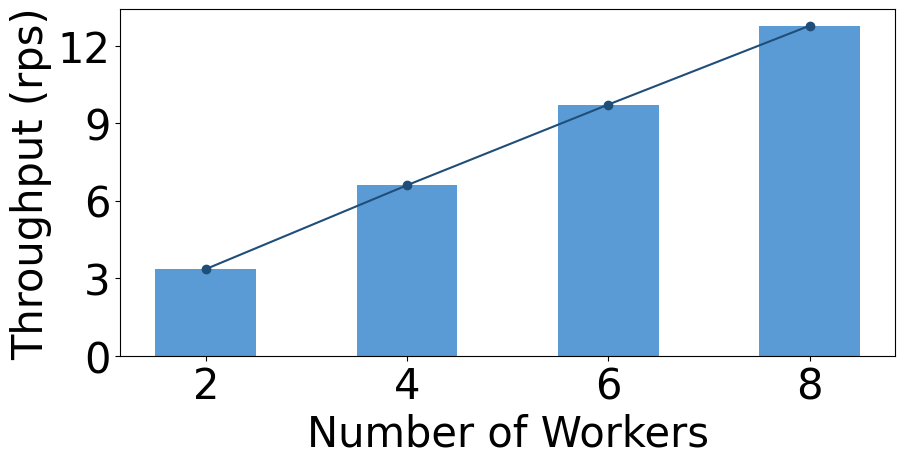}
		\caption{Deepspeed-Inference.}
		\label{fig:scale_ds}
	\end{subfigure}
	\caption{Scalability of SCLS.}
	\label{fig:scale}
	\vspace{-0.25cm}
\end{figure}

\subsection{Scalability}
We evaluate the throughput of SCLS under different numbers of workers at a request arrival rate of 20, where each worker is deployed on an NVIDIA A100 80GB GPU. The experimental results are shown in Fig. \ref{fig:scale}. We can see that the throughput grows linearly with the number of workers, which confirms the scalability of SCLS.

\section{Related Works}
\subsection{Model Compression}. 
Model quantization \cite{frantar2023optq, xiao2023smoothquant, shao2023omniquant, chee2024quip, lin2024awq} converts the numerical weight precision of LLMs into fewer bits to reduce the memory usage of LLM parameters. The technique of pruning \cite{xu2022dense, sun2023simple, ma2023llm, shao2024one} eliminates redundant components of LLMs to increase the inference speed. These approaches can affect LLM accuracy while SCLS is a request scheduling framework that does not make any modification to LLM parameters and does not affect LLM accuracy. Besides, since SCLS is an inter-instance request scheduling framework, these works are orthogonal to SCLS, and SCLS can also improve the serving efficiency for quantized and pruned LLMs.

\subsection{Low-level Optimization.} 
Kernel optimization techniques such as flash attention \cite{dao2022flashattention, dao2023flashattention} and paged attention \cite{kwon2023efficient} speed up LLM inference by implementing efficient CUDA kernels to improve computational efficiency.
Parallelism optimization \cite{huang2019gpipe, narayanan2019pipedream, narayanan2021efficient, miao2022galvatron, miao2023sdpipe} proposes to split the model at intra-layer and inter-layer granularity to enable distributed LLM inference. These studies are always employed to make a single LLM instance more efficient in inference. However, SCLS is a scheduling system that schedules requests to multiple workers. It leverages proposed adaptive batching and scheduling algorithms to fully use the inference engine and balance load among multiple workers. Moreover, these studies are orthogonal to SCLS as well. SCLS can be augmented with these low-level techniques.

\subsection{LLM Serving Systems.} 
Existing serving systems \cite{triton, tensorflowserving} adopt the SLS to serve requests in an FCFS manner with a fixed batch size. Orca \cite{yu2022orca} and Deepspeed-FastGen \cite{deepspeed-fastgen} leverage the ILS to improve the serving efficiency, but they adopt conservative memory management strategies that limit throughput. PiA \cite{zheng2024response} and $S^3$ \cite{jin2024s} propose to roughly predict the request generation length to enhance serving efficiency. However, they require significant computational overhead to fine-tune LLMs or train predictors.
Recent studies \cite{zhao2024atom, liu2023deja, fang2021turbotransformers, sheng2023flexgen, miao2024spotserve, zhong2024distserve, agrawal2024taming, fu2024serverlessllm, sheng2024fairness, sheng2024slora, wu2024dlora, abhyankarinfercept} propose a series of advanced techniques such as chunked prefill, disaggregated inference, and multi-lora serving. These works are designed for continuous batching, and we will combine these techniques with generation slicing in the future. 

\section{Generation Slicing for Continuous Batching}
\label{sec:discussion}
For continuous batching, if the generation length of a request is too long, it will consume a large amount of memory, which causes newly arrived requests to queue up and can not be added to the serving process due to insufficient memory. By limiting the number of generated tokens for requests in each schedule to a small slice length, we can solve this problem by timely re-scheduling the long request to an instance with more free memory. Besides, we can also accurately estimate the memory consumed by requests in each schedule to serve as many requests in parallel as possible without causing OOM errors or request preemption and achieve balanced memory consumption across multiple LLM instances. Besides, for requests re-scheduled to the same LLM instance, we can mitigate the prefill re-computation overhead by reusing the previously produced key-value cache with prefix-caching \cite{lin2024parrot} or proactively swapping the key-value cache between GPU and CPU memories. For requests re-scheduled to different LLM instances, the prefill re-computation can be amortized with decoding using chunked prefill \cite{agrawal2024taming}. 
\section{Conclusion and Future Works}
In this paper, we propose the core idea of generation slicing and design slice-level scheduling (SCLS) to achieve high-throughput and load-balanced LLM serving with static batching. By splitting the predefined maximal generation length limit into slices, SCLS can provide a precise range of serving time and memory usage for batches, laying the foundation for the design of adaptive bathing and max-min offloading algorithms. Extensive experiments are conducted to confirm the superiority of SCLS in throughput improvement and load balance. In the future, we will apply generation slicing to continuous batching and mitigate the overhead of the prefill re-computation by amortizing prefill computation using the chunked prefill technique and proactively swapping the key-value cache between the CPU and GPU memory.

\bibliographystyle{IEEEtran}
\bibliography{IEEEabrv,ref}

\vfill

\end{document}